\documentclass[sigconf, nonacm]{acmart}

\makeatletter
\def\@ACM@checkaffil{}
\makeatother

\usepackage{graphicx}
\usepackage{subcaption}

\newcommand{\mytab}{Tab.\@~}
\newcommand{\mysec}{Sec.\@~}
\newcommand{\myfig}{Fig.\@~}
\newcommand{\myapp}{App.\@~}

\newcommand{\myfiglong}{Figure~}

\newcommand{\mytablong}{Table~}

\begin{document}

\title{Llamas on the Web: Memory-Efficient, Performance-Portable, and Multi-Precision LLM Inference with WebGPU}

\author{Reese Levine}
\affiliation{%
  \institution{UC Santa Cruz}
}

\author{Rithik Sharma}
\affiliation{%
  \institution{UC Santa Cruz}
}

\author{Nikhil Jain}
\affiliation{%
  \institution{UC Santa Cruz}
}

\author{Abhijit Ramesh}
\affiliation{%
  \institution{UC Santa Cruz}
}

\author{Zheyuan Chen}
\affiliation{%
  \institution{UC Santa Cruz}
}

\author{Neha Abbas}
\affiliation{%
  \institution{UC Santa Cruz}
}

\author{James Contini}
\affiliation{%
  \institution{UC Santa Cruz}
}

\author{Tyler Sorensen}
\affiliation{%
  \institution{Microsoft Research}
  \institution{UC Santa Cruz}
}

\begin{abstract}
Running language models in the browser presents a unique opportunity to build efficient, private, and portable AI applications, but requires contending with constrained memory availability and heterogeneous hardware targets. To realize this opportunity, we present Llamas on the Web (LlamaWeb), a WebGPU backend for llama.cpp that enables memory-efficient and performance-portable LLM inference across a wide range of model weight formats in the browser. Our design significantly reduces memory overhead through static memory planning and efficient model loading, addresses cross-device variability through a tunable kernel library, and introduces templated GPU kernels that support performant implementations of numerous quantization formats, enabling broad model support and extensibility to new formats.


We evaluate LlamaWeb on 16 devices from 8 vendors, collecting data from 10 language models and four model weight formats. We compare LlamaWeb against existing browser-based LLM frameworks and find that LlamaWeb requires 29--33\% less memory across several combinations of device, browser, and operating system. We also evaluate LlamaWeb's performance against these frameworks and find that it increases decode throughput by 45--69\% across four GPUs from separate vendors. In addition, we compare LlamaWeb's performance against other llama.cpp backends, where it is competitive with and even beats vendor-specific backend performance on some devices.


\end{abstract}

\maketitle

\begin{table}
\small
\caption{Comparison of LlamaWeb (this work) to other browser-based LLM inference frameworks on key metrics.}
\label{tab:cool-info}
\begin{tabular}{l r r r}
\toprule
& \textbf{LlamaWeb} & \textbf{WebLLM} & \textbf{Transformers.js} \\
\cmidrule(lr){2-4}
\textbf{Available Models}$^{1}$ 
    & \textbf{177,691} 
    & 400 
    & 41,632 \\

\textbf{Memory Usage}$^{2}$ 
    & \textbf{Baseline}
    & +49\%
    & +41\% \\

\textbf{Performance}$^{3}$ 
    & \textbf{Baseline}
    & -35\%
    & -41\% \\

\textbf{Weight Formats}$^{4}$ 
    & \textbf{23}
    & 6
    & 7 \\
\bottomrule
\end{tabular}
{\footnotesize
\raggedright
$^{1}$ Number of compatible models available in May 2026, measured using publicly accessible Hugging Face repositories (\myapp\ref{app:model-support}). \\

$^{2}$ Geometric mean comparison of normalized peak browser memory usage across several configurations during inference of an \texttt{f16} Llama3.2 1B model (\mysec\ref{sec:eval-mem}). \\

$^{3}$ Geometric mean comparison of normalized decode tokens per second of the same Llama3.2 model across 4 GPUs (\mysec\ref{sec:eval-perf-port}). \\

$^{4}$ Number of supported floating-point and quantization formats available for inference in WebGPU (\myapp\ref{app:quant-support}). \\
}

\end{table}

\section{Introduction}
\label{sec:intro}

The capabilities of (smaller) large language models (LLMs) are improving rapidly~\cite{saadfalcon2026intelperwatt, belcak2025smalllanguagemodelsfuture}, with many model developers releasing open-weight models that are explicitly designed for edge deployments~\cite{bonsai2025, gemmateam2025gemma3technicalreport, allal2025smollm2smolgoesbig}. Running models locally can improve inference latency and energy efficiency~\cite{husom2025edgeai}, especially as AI datacenter power usage continues to grow~\cite{ai_index_2026, jegham2025hungryaibenchmarkingenergy}. Additionally, local models provide privacy benefits, with a recent study finding that many frontier AI companies retain and utilize user conversations for training by default and lack transparent privacy policies~\cite{king2025userprivacylargelanguage}.

Capitalizing on these benefits, inference engines designed to run on consumer and edge devices like llama.cpp~\cite{llamacpp} and MLX~\cite{mlx2023} have risen in popularity over the past few years. These engines often ship with vendor-specific GPU backends, e.g., MLX is optimized for Apple GPUs, require expert knowledge to setup correctly, and must be installed separately on a user's system, e.g., on the command line or through an app. On the other hand, web browsers present an attractive choice for local LLM deployments due to their integration into many users' daily tasks~\cite{zhou2024webarenarealisticwebenvironment} as well as their ease of use and lack of external software dependencies. New frameworks like WebGPU~\cite{webgpu} enable GPU acceleration within the browser, opening up opportunities for interactive, low-latency usage of language models. Several existing frameworks such as Transformers.js~\cite{transformers_js_docs}, WebLLM~\cite{ruan2026webllmhighperformanceinbrowserllm}, and wllama~\cite{wllama} target browser-based inference, with Transformers.js and WebLLM supporting GPU acceleration through ONNX Runtime~\cite{onnxruntime_web_docs} and MLC-LLM~\cite{mlc-llm}, respectively.

However, existing browser-based systems are limited by several issues:
\begin{itemize}
    \item \textit{Memory Inefficiencies}: Existing browser inference engines dynamically allocate GPU memory and load weights into WebGPU inefficiently, leading to slowdowns as memory grows over the course of execution and even crashes as some browsers enforce hard-caps on memory per-tab.
    \item \textit{Lack of Performance-Portable Design:} Existing frameworks have limited kernel specialization for different devices, a critical limitation as WebGPU provides \textit{functional} portability but not performance portability. Maximizing performance on the diverse hardware targeted by browsers motivates the need for interfaces that allow kernels to be tuned independently and easily integrated into inference engines.
    \item \textit{Limited Quantization Support:} To fully realize the opportunity of running models locally, inference engines must support diverse quantization formats that allow upstream model developers to experiment with strategies like quantization aware training and dynamic weight quantization. Kernels that operate on quantized models must be performant while allowing new formats to be added as needed.
\end{itemize}

Taken together, these limitations hinder the adoption of cross-platform and browser-based LLM inference, motivating the need for new research into techniques to overcome them. 


\subsection{LlamaWeb}
\label{sec:intro-llamaweb}

In this work, we introduce LlamaWeb, a WebGPU backend for llama.cpp that runs many models supported by llama.cpp with GPU acceleration in the browser. Our approach is guided by three key constraints: (1) memory usage must be minimal and statically allocated, reducing overhead and allowing useful models to run in the browser on edge devices while avoiding unexpected slowdowns and crashes, (2) kernel libraries must be flexible and extensible to support performance portability across WebGPU's heterogeneous hardware targets, and (3) support for diverse quantization formats needs to be a first-class concern.

By building within llama.cpp, LlamaWeb also benefits from an active open-source ecosystem; for example, the first row in \mytab~\ref{tab:cool-info} shows the number of models available in llama.cpp's GGUF format on Hugging Face, as compared to models available in other WebGPU frameworks' formats.  In turn, our work extends this ecosystem to the browser, enabling the broad collection of models, quantization strategies, and runtime features developed for llama.cpp to execute efficiently on WebGPU-enabled devices without requiring platform-specific native deployments.

\paragraph{Memory-Efficient Local Inference}

A central design goal of our system is minimizing memory overhead during inference. To this end, we statically allocate all memory necessary to run the model at startup, including intermediate data structures for operations like FlashAttention and a fixed-size slotted memory region used for kernel parameters. We also optimize model loading and data movement, avoiding unnecessary materialization of model weights and streaming them asynchronously to GPU memory.
In \mysec\ref{sec:eval-mem}, we compare memory consumption across multiple devices and two browsers (Chrome and Safari) and show that these strategies reduce peak memory usage by 29--33\% on average compared to existing browser-based inference frameworks.

\paragraph{Performance-Portable Kernel Library}

Maximizing performance across diverse GPU architectures requires adaptable kernel implementations. We develop a kernel library enabling templated shader generation with device-specific optimization and performance-portable parameters. For example, parameters like tile sizes for matrix multiplication can be specialized for different devices or use performance-portable values. Kernels can also incorporate advanced features like subgroup operations or subgroup matrix instructions when supported, while falling back to more portable implementations otherwise. At the execution level, we implement efficient grouping of kernel submission to the GPU and avoid unnecessary CPU-GPU synchronization. 

In \mysec~\ref{sec:eval-perf-port}, we evaluate this system across a diverse set of models and devices, running 10 models across 16 devices from 8 vendors. To our knowledge, this is the largest cross-device, cross-model evaluation dataset collected within a single inference engine framework, and is the only WebGPU dataset that includes mobile devices. We compare our performance against native backends, showing how our kernels achieve portable performance and even beating the performance of native backends in some instances. We further compare our performance against WebLLM and Transformers.js, with LlamaWeb outperforming these frameworks by 54--69\% on average during decode. However, during prefill, LlamaWeb underperforms these frameworks by 21--51\% during prefill in the browser, which highlights room for future improvement.

\paragraph{Quantization-Aware Shader Design}

Supporting the wide variety of quantization formats used in llama.cpp requires careful co-design of kernels and quantized data layouts. Therefore, we develop templated kernels that integrate various dequantization routines directly into computation, e.g., by dequantizing into shared memory or registers while performing row-column reductions. Our approach allows us to implement many of llama.cpp's model weight formats and far more than other frameworks, as shown in the last row of \mytab~\ref{tab:cool-info}. Our routines are also reusable; for example, the same logic is used when dequantizing weights for matrix multiplication and when accessing KV-cache entries during FlashAttention. To evaluate the effectiveness of our design, we evaluate a Llama3.2 model across multiple quantization levels on a range of devices (\mysec\ref{sec:eval-quant}), analyzing performance and efficiency trends. 

\paragraph{Contributions} In summary, our contributions are:
\begin{itemize}
    \item A memory-efficient inference engine implementation in WebGPU that maximizes the number of models that can be run on memory-constrained devices and maintaining performance and stability across various platforms (\mysec~\ref{sec:llamaweb-mem}).
    \item A flexible and extensible shader library for LLM inference that allows for external tuning of important parameters, specializes shaders based on available features, and is powered by an optimized scheduling runtime (\mysec~\ref{sec:llamaweb-kernel}).
    \item A set of core LLM kernels, namely matrix operations, that are carefully co-designed with llama.cpp's numerous quantization formats, enabling the deployment of models optimized for different sizes and allowing easy integration of emerging quantization methods (\mysec~\ref{sec:llamaweb-quant}).
    
\end{itemize}

Our core WebGPU backend, which currently consists of 8,470 lines of C++ and 44 kernel template files with 11,338 lines of code, supports 72 core ML operators and 23 data formats, including 32 and 16-bit floating point types and 21 quantization formats. All our code is open-source and has been integrated into llama.cpp codebases through over 100 pull requests, and we look forward to seeing how the community builds on our work going forward.

\section{Background}
\label{sec:background}

\subsection{Language Model Inference}
\label{sec:background-llm}

Modern large language models (LLMs) are predominantly auto-regressive, where tokens, ranging from a single character to an entire word, are sequentially generated conditioned on previously generated outputs. The main class of such models are transformer-based architectures ~\cite{vaswani2017attention}, which rely on the self-attention mechanism to process and generate new tokens. Alternative approaches have been proposed to address the quadratic increase in computation and memory as sequence length increases in attention mechanisms, including architectures based on state space models~\cite{gu2024mambalineartimesequencemodeling} and gated convolutions~\cite{amini2025lfm2technicalreport}.

Most modern models' inference runtime is dominated by linear algebra operations such as matrix–matrix and matrix–vector multiplications. This runtime can be broken down into two distinct phases: (1) \textit{prefill}, where an entire prompt is processed, and (2) \textit{decode}, where new tokens are generated sequentially. During prefill, dense matrix multiplication is used to process all tokens in the prompt, while during decode matrix-vector multiplication is often used to process a single token. A model's representation of previously processed tokens in the attention mechanism can be stored in a \textit{KV-cache}, which is utilized to compute attention scores when generating new tokens. The complexity of attention has led to the development of hardware-efficient mechanisms like FlashAttention~\cite{dao2022flashattention}, which fuses multiple matrix multiplications with a softmax normalization to avoid multiple round-trips to DRAM on GPUs.

Inference engines such as llama.cpp~\cite{llamacpp} provide a concrete implementation of these ideas. In llama.cpp, model weights are stored in a custom binary format (GGUF) which encodes tensors containing model weights, along with metadata, and can be distributed through model repositories like Hugging Face. The model architectures themselves are described within the llama.cpp codebase. At execution time, a dynamic graph of tensor operations is constructed, ranging from metadata-only operations that do not access or move the underlying data, e.g., reshape and transpose, to materializing operations that do move or compute on tensor data, e.g., matrix multiplication. This directed acyclic graph (DAG) is then linearized into a topological ordering and executed by different backends. 

The llama.cpp codebase contains various backends that allow models to be executed on diverse hardware, including CPUs, GPUs, and NPUs. Most existing backends target devices from specific vendors, including a CUDA backend for NVIDIA GPUs, a Metal backend for Apple GPUs, and a HIP backend for AMD GPUs. Cross-platform backends like Vulkan and OpenCL also exist, but no existing backends were designed with the constraints of browser-based environments in mind.

Some inference engines, such as vLLM~\cite{kwon2023vllm}, are specifically designed for multi-GPU and server-based inference. These engines incorporate sophisticated sequence-batching strategies that maximize throughput and avoid GPU under-utilization during the decode phase. While llama.cpp is evolving to support these use-cases as well, our WebGPU implementation is specifically focused on low-latency single-GPU inference, e.g., in the browser, so we do not explore batching in this work.

\subsection{Model Quantization}
\label{background:quant}

The size of modern language models presents a major barrier to local deployment. Storing model weights in full 32-bit or even 16-bit floating point format leads to large memory requirements, often exceeding the capacity of consumer devices. As a result, quantization techniques have become essential for enabling efficient inference. Quantization reduces the precision of model weights (and sometimes activations) to lower-bit representations, such as 8-bit or 4-bit integers, reducing both memory footprint and bandwidth requirements. Llama.cpp supports a variety of quantization formats, with most following the form:

\begin{equation}
q_i = \mathrm{round}\!\left(\frac{x_i - \mu}{s}\right),
\qquad
\hat{x}_i = s \cdot q_i + \mu
\label{eq:quant}
\end{equation}

where $x_i$ is the original weight, $s$ is a scaling factor, $\mu$ is an optional offset depending on the format, $q_i$ is the computed quantized value, and $\hat{x}_i$ is the resulting dequantized weight. Weights are quantized by grouping them into blocks and choosing scaling factors based on the range of weight values within that block, e.g., by calculating the absolute maximum. The three major quantization formats supported by llama.cpp are:

\begin{itemize}
    \item \textbf{Basic (legacy):} These formats use blocks of 32 weights. Variants like \texttt{q4\_0} and \texttt{q8\_0} are \textit{symmetric}, i.e., they do not include offsets, while variants like \texttt{q4\_1} compute and store offsets per-block.
    \item \textbf{K-quants:} These formats use blocks of 256 weights, which are subdivided into blocks of 32, each symmetric and with their own scaling factor. Scaling factors are themselves quantized into a single super-block scaling factor, further reducing the memory overhead. While compression of quantization constants has a long history in signal processing~\cite{gersho1991quant}, the design decisions behind llama.cpp K-quants were never explicitly laid out~\cite{kawrakow_kquants_2023}. However, they are similar to the double quantization technique introduced in QLoRA~\cite{dettmers2023qlora}.
    \item \textbf{I-quants:} These formats also use blocks of 256 weights, but are inspired by vector quantization, including methods like QuIP\#~\cite{tseng2024quip}, where small groups of weights (usually 8) are encoded as an index to a codebook of reference vectors.
\end{itemize}

All of these quantization formats are \textit{generic}; that is, they are not designed for particular hardware, but rather dequantization is implemented in software and operations are performed on the resulting values. Along with these generic formats, llama.cpp also supports hardware-native formats like \texttt{bf16}, \texttt{nvp4}, and \texttt{mxfp4}. These formats can also be emulated in software, and the WebGPU backend does support \texttt{mxfp4} emulation in order to run OpenAI's gpt-oss-20b model~\cite{openai2025gptoss}, but emulation for other native formats is not currently supported. Recently, a quantization format called \texttt{q1\_0} was introduced to support the Bonsai 1-bit model family~\cite{bonsai2025}, which uses symmetric quantization with a single scale for a block of 128 weights. Due to our quantization-aware kernel design, we were able to easily integrate support for this quantization type with minimal effort, as described in \mysec~\ref{sec:llamaweb-quant}.

Despite their widespread use, there is limited formal documentation or literature describing the design and especially the data layout of these quantization formats. Many formats rely on non-contiguous or highly specialized packing schemes that may have been optimized for specific hardware access patterns. 
Understanding the trade-offs in quantization block layout with respect to vectorization, memory alignment, and GPU execution remains an open area for future work.

The quantization formats in llama.cpp are \textit{weight-only}. However, llama.cpp has also incorporated techniques from other post-training quantization methods like GPTQ~\cite{frantar2023gptq} and AWQ~\cite{lin2026awq} into its quantization tools, and the llama.cpp community is also working on novel techniques for mixed-precision quantization using llama.cpp's standard formats. Model developers can also perform quantization-aware training~\cite{jacob2017qat} using a given quantization format and deploy the resulting model with llama.cpp.


\subsection{WebGPU Programming Model}
\label{sec:background-webgpu}

WebGPU is a modern cross-platform graphics and compute API designed to provide low-level access to GPU hardware within web browsers. Despite its name, WebGPU can also be used directly by native applications, as some WebGPU implementations are independent libraries that can run outside browser contexts. Like most other GPU programming models, WebGPU separates host-side control logic from device-side shader execution. While WebGPU supports both graphics and compute use-cases, for this work we focus only on its compute capabilities.

On the device side, kernels (called \textit{shaders} in WebGPU) are written in the WebGPU Shading Language (WGSL), which supports general-purpose compute in a Single Instruction, Multiple Threads (SIMT) execution model. Threads are organized into workgroups, where each thread has access to private registers, workgroups share fast on-chip memory (analogous to shared memory in CUDA), and all threads can access global device memory. Recent extensions to the WebGPU programming model include subgroups (analogous to warps in CUDA), which represent smaller collections of threads that can execute cooperative operations such as reductions or shuffles using registers. Emerging features such as subgroup matrix operations further expose specialized hardware units, e.g., tensor cores, for efficient matrix multiplication.

The WGSL language supports various data-types, including 32 and 16-bit floating points (\texttt{f32} and \texttt{f16}), signed and unsigned 32-bit integer types (\texttt{i32} and \texttt{u32}), and vector types. It does not yet support specialized types like \texttt{bf16}~\cite{google_bfloat16_blog_2019}, smaller integer types like \texttt{u16} and \texttt{u8}, or specialized 4-bit types like \texttt{nvfp4}~\cite{nvidia_nvfp4_blog_2025} and \texttt{mxfp4}~\cite{ocp_mx_spec_2023}.

On the host side, applications interact with the GPU (\textit{device}) through a structured API. Kernels are compiled at runtime into \textit{pipelines}, which define both the executable code and the layout of bound resources. Data is stored in \textit{buffers} allocated by the host and organized into \textit{bind groups} which are bound to pipelines according to specified \textit{bind group layouts}. WebGPU follows a deferred execution model; compute workloads are recorded via \textit{command encoders} and \textit{compute passes}, but this work does not execute on the GPU until the command encoder is \textit{finished}, i.e., flushed, into a \textit{command buffer}, which is in turn submitted to the device \textit{queue}. Along with command buffers, work like copying data from one buffer to another can be submitted to the queue. The granularity of queue submission, such as how many operations are grouped into a single compute pass or how many commands are in flight in the queue, can affect performance and stability, and application performance can vary across different WebGPU implementations.

WebGPU is designed as a portability layer over multiple native APIs, with current implementations supporting Metal on macOS/iOS~\cite{apple_metal}, Vulkan on Linux and Android~\cite{vulkan1.3}, and DirectX on Windows~\cite{directx-specs}. Many major browsers have their own WebGPU implementations: Dawn~\cite{dawn} is used in Chromium-based browsers, wgpu~\cite{wgpu} in Firefox, and WebKit~\cite{WebKit} provides its own implementation for Safari. While WebGPU is primarily designed for browsers, implementations like Dawn, which is written in C++, can also be used with native code. Tools like Emscripten~\cite{zakai2011emscripten} enable cross-compilation of C++ to WebAssembly (WASM), a portable low-level bytecode that, like WebGPU, is built to abstract over different hardware and platforms~\cite{haas2017wasm}. Therefore, although we mainly focus on inference in the browser in this work, LlamaWeb can also be used in any environment that supports C++ or WASM.


\begin{figure}
\includegraphics[width=\columnwidth]{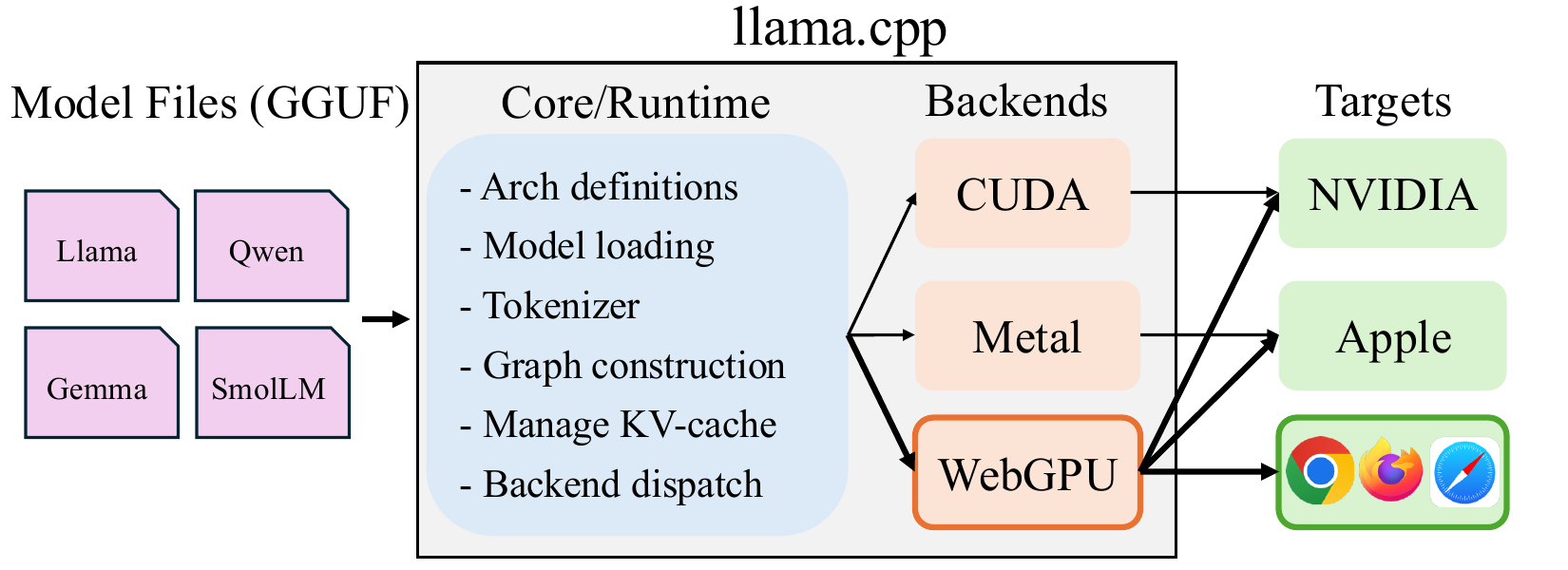}
\caption{Overview of llama.cpp's core and backend design.}
\label{fig:overview}
\end{figure}

\begin{figure}
\includegraphics[width=\columnwidth]{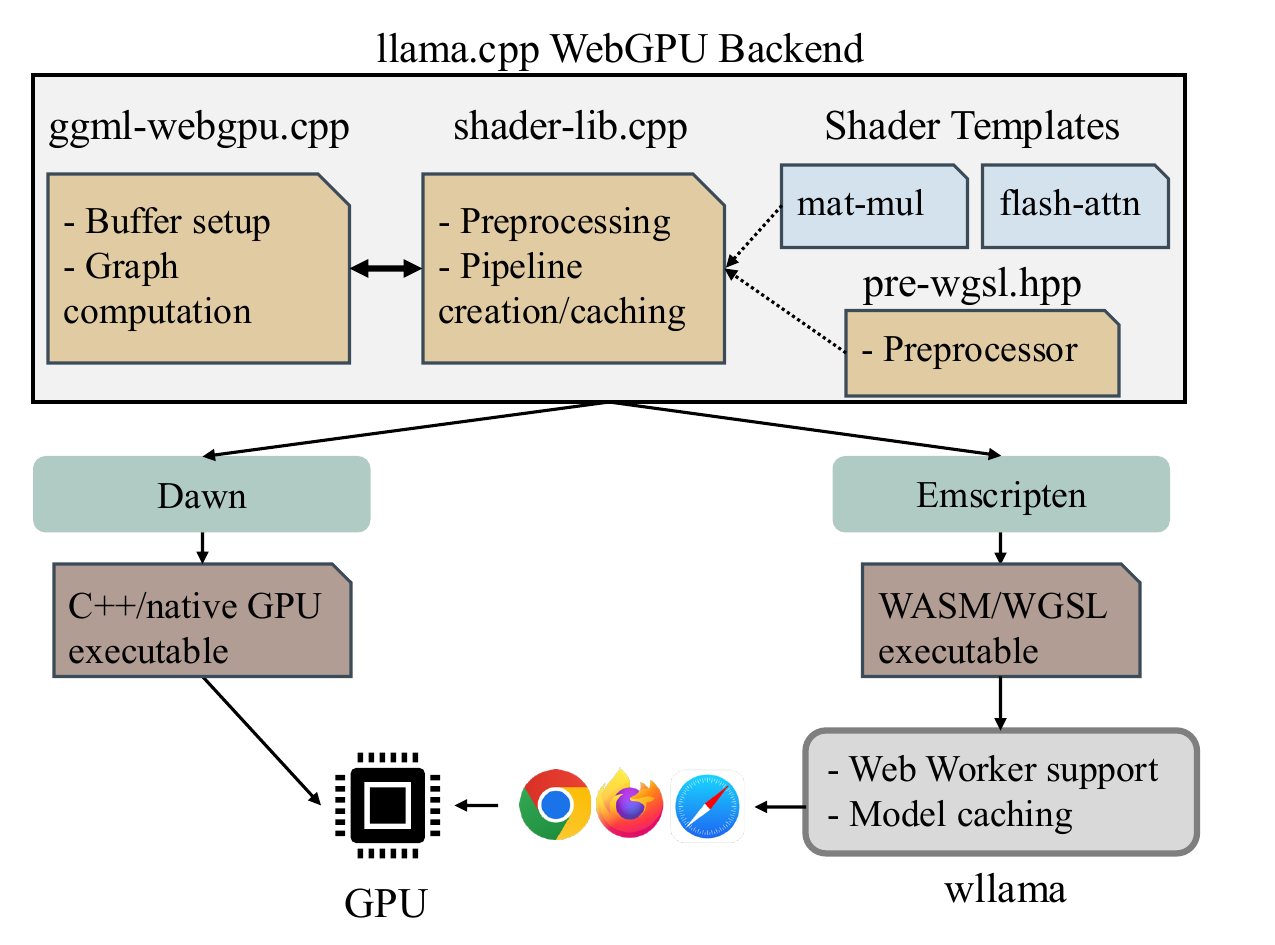}
\caption{Breakdown of the LlamaWeb llama.cpp WebGPU backend and its different paths for executing on GPUs. The backend can be built for native execution using Dawn as the WebGPU implementation, or built for cross-browser execution with Emscripten.}
\label{fig:webgpu-detail}
\end{figure}

\section{Design of LlamaWeb}
\label{sec:llamaweb}

\myfiglong\ref{fig:overview} gives an overview of the llama.cpp inference engine. GGUF files are loaded by llama.cpp and the tensor operation graph is dispatched to backends through a standardized interface. As mentioned, llama.cpp supports numerous backends; our work, the highlighted WebGPU backend, targets GPUs from many vendors and adds support for GPU-accelerated inference in the browser.

\myfiglong\ref{fig:webgpu-detail} shows a more detailed diagram of how the WebGPU backend is implemented. The main implementation is split into two files: \texttt{ggml-webgpu.cpp}, which implements llama.cpp's backend interface, and \texttt{shader-lib.cpp}\footnote{In this paper we use the term kernels to refer to shaders, but in our code we follow the WebGPU convention of calling them shaders.}, which handles preprocessing and compilation of kernels into WebGPU pipelines. 

As WebGPU does not have a built-in preprocessor, many WebGPU applications have developed ad-hoc preprocessing, e.g., using JavaScript string substitution and concatenation. Motivated by our need for a more standard solution that works within llama.cpp's C++ codebase and is familiar to developers, we built a minimal WGSL preprocessor which we call pre-wgsl~\cite{pre-wgsl}. Pre-wgsl is a header-only library that is directly included in the WebGPU backend and can also be used independently either natively or in web environments.

To run llama.cpp with WebGPU acceleration in the browser, we add WebGPU support to the pre-existing wllama library, which previously only supported running llama.cpp through WASM on the CPU. Wllama includes features like caching models in the browser's Origin Private File System (OPFS)~\cite{opfs} and running WebGPU in a Web Worker to avoid freezing the UI thread. In the rest of this section, we describe in more detail the key design points of LlamaWeb. 

\subsection{Memory-Efficient Browser Inference}
\label{sec:llamaweb-mem}



In LlamaWeb, memory is statically allocated at startup, ensuring predictable memory usage independent of prompt or generated output size. To accomplish this, we do the following: (1) implement a static parameter buffer ``arena'' which avoids dynamic GPU buffer creation and complicated caching logic, (2) calculate up-front any extra memory needed for intermediate states, e.g., to implement FlashDecoding~\cite{flashdecoding2023}, and (3) optimize model loading in the browser to avoid redundant memory and large CPU memory allocations. Inherited from llama.cpp, LlamaWeb also utilizes a fixed-size KV-cache which can be allocated before running inference.

Each kernel needs a small number of dynamic parameters, e.g., matrix dimensions, passed in. Unlike in other languages, e.g., using push constants in Vulkan, WebGPU does not yet support a way to directly pass small amounts of data to kernels at runtime, so frameworks like WebLLM and ONNX Runtime maintain pools of small buffers that are dynamically allocated and cached. Though the memory overhead of these pools is small, when pushing against the limits of a device's memory, allocating a new buffer may be the difference between crashing the page or not. Therefore, LlamaWeb allocates a single buffer at startup with enough slots to fit the parameters for a configurable number of kernels. Within the buffer, we rotate through the slots, and we guarantee that parameters are not overwritten until the kernel that relies on them finishes. This simple design also avoids any synchronization logic needed to maintain a buffer pool, including handling asynchronous callbacks to release buffers, and avoids memory fragmentation.

During some operations, e.g., FlashDecoding, several workgroups cooperate on computing attention scores across a single query vector, and per-workgroup results are stored in an intermediate buffer which is reduced by a separate kernel. Also for FlashDecoding, as well as for FlashAttention and sampling operations like \texttt{top-k}, we implement optimizations that run several kernels, some of which require intermediate memory storage space. In all cases, LlamaWeb allocates this extra memory \textit{before} the model first runs.

Lastly, we optimize wllama to efficiently load model weights. First, we avoid redundant memory copies by downloading and storing weights in disk (utilizing OPFS) and never materialize them in the WebAssembly heap\footnote{This is especially important because the WASM heap is grow-only~\cite{watt2019weakwasm}, so once memory is allocated by a tab it can't be released during the tab's lifetime.}. Next, we utilize llama.cpp's asynchronous loading interface to reduce the memory footprint of the loading process, using only four 1 MB buffers to stream weights directly from OPFS into WebGPU buffers. In particular, we found that Safari has especially strict memory usage limits, so these changes enabled us to serve larger, more powerful models with our system.

\begin{table}[t]
\small
\caption{Percentage of time spent in different kernel categories during prefill (512 token prompt) and decode (128 tokens generated) when running LlaMA 3.2 1B (\texttt{q4\_k\_m} format) on an Apple M3 at KV-cache depths of 0 and 2048.}
\label{tab:webgpu-kernel-breakdown}
\centering
\begin{tabular}{l r r r r}
\toprule
& \multicolumn{2}{c}{\textbf{KV@0}} & \multicolumn{2}{c}{\textbf{KV@2048}} \\
\cmidrule(lr){2-3} \cmidrule(lr){4-5}
\textbf{Kernel Category} & \textbf{prefill} & \textbf{decode} & \textbf{prefill} & \textbf{decode} \\
\midrule
Matrix multiplication & 92.6\% & 0.0\% & 83.4\% & 0.0\% \\
Matrix-vector multiplication & 0.5\% & 89.9\% & 0.3\% & 80.7\% \\
Attention & 4.4\% & 5.3\% & 14.1\% & 15.0\% \\
Normalization/elementwise & 2.2\% & 3.0\% & 2.0\% & 2.7\% \\
Other & 0.3\% & 1.8\% & 0.3\% & 1.6\% \\
\bottomrule
\end{tabular}
\end{table}

\subsection{Kernel Library and Execution Scheduling}
\label{sec:llamaweb-kernel}

The WebGPU backend is separated into two layers: the runtime execution logic and the kernel library. The runtime is responsible for high-level orchestration; it constructs and passes a lightweight context to the library that describes the tensors involved, e.g., shapes, strides, and data types, along with device-specific information such as limits and supported features. Then, given compiled pipelines returned by the kernel library, the runtime handles logic for execution scheduling and grouping operations into compute passes, managing command buffer submission, and coordinating CPU–GPU synchronization.

Given an operation and context, the kernel library first performs a preprocessing step to construct an appropriate kernel variant. This includes decisions such as selecting code blocks based on tensor data formats (e.g., \texttt{f32}, \texttt{f16}, or quantized formats), choosing between workgroup-level or subgroup-level reductions depending on device features, determining whether an operation must be performed in-place (which affects how buffers are bound to the kernel as WebGPU does not allow buffer aliasing), and configuring tiling or vectorization strategies. Based on these choices, the library generates kernel code and compiles it into a GPU pipeline. To avoid redundant work, compiled pipelines are cached using a key that encodes the information used to specialize the kernel. Alongside the compiled pipeline, the kernel library can also return metadata describing the decisions it made, which can inform subsequent dispatch shapes and scheduling decisions. 

This separation of concerns between the kernel library and runtime execution logic means that in the future, the library can independently evolve to support increasingly sophisticated performance-portable strategies, including device-specific tuning, dynamic specialization, and even auto-tuning mechanisms. Currently, we choose performance-portable parameters for operations like matrix multiplication based on a large empirical study, while other operations use heuristically chosen values for parameters like workgroup size based on pilot experiments. Kernel compilation incurs a one-time cost (\textasciitilde 1-5 seconds) during the first forward pass on the model, after which execution proceeds using cached pipelines.

Several categories of kernels are necessary to run a full forward-pass on a model. \mytablong~\ref{tab:webgpu-kernel-breakdown} shows the percentage of time spent executing each kernel category for a representative benchmarking run. Matrix and matrix-vector multiplication kernels dominate runtime in prefill and decode respectively, while normalization/elementwise and other operations, including rotary position encoding and tensor data movement, are a small percentage of runtime. As expected, time spent doing attention grows as the KV-cache depth increases. 

We now briefly describe our implementations of these different kernel categories. All kernels are validated using llama.cpp's testing infrastructure, which compares the output of running the operation on the GPU to running a reference implementation on the CPU, and ensuring the numerical mean squared error (NMSE) across the output remains under a threshold, commonly $1e^{-7}$. During development, we observed that differences in floating-point rounding behavior while casting from \texttt{f32} to \texttt{f16} values caused higher NMSE on some devices, e.g., an NVIDIA Tesla T4, so for operations on \texttt{f16} tensor data the NMSE threshold is relaxed to $1e^{-6}$.

\paragraph{General Purpose Kernels} Our library includes kernels performing a wide variety of general-purpose tasks on the GPU. For example, it supports many unary, e.g., \texttt{floor}, \texttt{ceil}, \texttt{abs}, and binary, e.g.,
\texttt{add}, \texttt{mul}, \texttt{div}, operations. It also includes support for important LLM operations like gated linear units (\texttt{glu}), rotary position embedding (\texttt{rope}), and normalizations, and for sampling\footnote{We find that offloading single-token sampling to the GPU in WebGPU currently does not improve performance. Optimization here is an area for future work.}, operations like \texttt{top-k} and \texttt{argmax}. To support state space models, the library also includes support for convolutions and GPU prefix scans. Although general operations like these are normally not the largest bottleneck in LLM inference, our library design allows these kernels to be optimized and specialized as necessary, e.g., by using subgroup reductions or vectorized memory accesses.

\paragraph{Matrix Multiplication} As it is one of the most important operations used in LLM inference, the library implements extensive support for matrix multiplication in \texttt{f32}, \texttt{f16}, and quantized formats. It includes two versions of matrix-matrix multiplication, \texttt{reg\_tile} and \texttt{sg\_mat}, which are primarily used during the prefill phase of LLM inference. Both follow a hierarchical tiling strategy; \texttt{reg\_tile} first tiles into shared memory, then into registers per-thread\footnote{During development we observed that using \texttt{f16} accumulation for the register-tiling kernel caused some models, e.g., Qwen2.5, to generate incoherent output on Apple M-series GPUs, so we currently use \texttt{f32} accumulation for this kernel. Differences in floating-point precision is a well-known issue, and handling it within WebGPU and its applications is not a solved problem.}, while \texttt{sg\_mat} also tiles into shared memory, but then utilizes WebGPU's newer subgroup matrix feature to run cooperative matrix multiplications on specialized hardware like NVIDIA's tensor cores. The subgroup matrix feature is not yet available in stable browser releases, but can be used natively in some WebGPU implementations.

The library also includes a specialized matrix-vector multiplication kernel for the decode phase. This kernel tiles the vector in registers and does a cooperative reduction across each row. When subgroups are available, the kernel is specialized to perform a subgroup reduction, otherwise falling back to a generic shared memory reduction. Each of these kernels exposes tunable workgroup sizes, shared-memory tile sizes, and per-thread or per-subgroup tile sizes. 

To select performance-portable defaults, we ran a large empirical study, exhaustively sweeping thousands of configurations across four GPUs from four different vendors over matrix shapes drawn from representative LLM workloads. From this data we derived performance-portable defaults that maximize average performance across the GPUs while minimizing worst-case slowdowns, delivering a 41\% kernel-level speedup on average over hand-picked values chosen during development (which was primarily done on systems with Apple GPUs). 
Pushing this data-driven approach further to more devices, more workload sizes, and more browser configurations is a natural direction for future work (Sec.~\ref{sec:future}).

\paragraph{FlashAttention}
As attention is one of the central operations in LLM inference, our library implements several variants. Rather than materializing the intermediate \texttt{QK\textasciicircum T} scores and softmax probabilities, these kernels stream over the KV cache in tiles and maintain the online-softmax state directly inside the kernel. The FlashDecoding implementation is optimized for decode; it maps one query row to one workgroup, keeps the row maximum, exponential sum, and output accumulator local to that row, and iterates over cached K/V tiles. The \texttt{tile} path targets larger query chunks, e.g., during prefill, by processing multiple query rows per workgroup and staging Q/K/V tiles in workgroup memory for better reuse. The implementation also contains a subgroup-matrix variant for environments where this experimental WebGPU feature is supported.
The backend also supports quantized KV-cache formats such as \texttt{q4\_0} and \texttt{q8\_0} by dequantizing K/V blocks while loading them from global to shared memory in the attention kernel. 

\subsection{Quantization-Aware Kernel Design}
\label{sec:llamaweb-quant}

As introduced previously, llama.cpp supports a diverse and evolving set of quantization formats, ranging from legacy schemes, e.g., \texttt{q4\_0}, \texttt{q8\_0}, to more advanced formats such as K-quants, I-quants, and most recently \texttt{q1\_0}. Dequantization is primarily needed in operations involving model weights, e.g., matrix-matrix and matrix-vector multiplications. In these operations, weights are stored in compressed formats and must be dequantized on-the-fly during computation. However, the activations in the KV-cache used in attention can also be stored in quantized form, requiring dequantization during attention score computation. 

A central challenge arises from the diversity of quantization layouts. Formats differ in block sizes, scaling factors, and bit-packing strategies. WebGPU has limitations on how types like structs are loaded from memory, as well as the required alignment of data types. To achieve better performance and portability, in the kernels all quantization formats are represented as flat buffers of unsigned 32-bit (\texttt{u32}) types, with dequantization routines interpreting these values depending on a format specified at compile time\footnote{For efficient memory loading, it would be more beneficial to represent some quantization formats as buffers of \texttt{u16}s, but WebGPU does not yet support this type.}.

Dequantization routines are integrated directly into kernels, with the particular routine chosen during preprocessing by the kernel library at compile time. For matrix multiplication, which is primarily compute-bound on most GPUs, we adopt a shared-memory tiling strategy; threads within a workgroup collaboratively load quantized blocks, dequantize them into shared memory, and reuse the decoded values across multiple output elements. Beyond weight operations, these routines can also be reused for KV-cache dequantization during attention. 

In contrast, matrix-vector multiplication is a memory-bound operation with limited data reuse. During implementation of this operation, we determined that a shared-memory tiling strategy for matrix-vector multiplication, as appears in other cross-platform portable libraries like CLBlast~\cite{Nugteren_2018}, was not a good fit for the many dequantization routines we wanted to support. Instead, we implement routines that dequantize directly into registers, improving access patterns to quantized memory and allowing the same underlying kernel to support llama.cpp's diverse quantization families. 

Our implementation supports the majority of quantization formats available in llama.cpp, including legacy formats, K-quants, I-quants, and \texttt{q1\_0}. Notably, this support is achieved without requiring format-specific kernel rewrites and enables rapid extensibility: for example, when \texttt{q1\_0} was introduced alongside the Bonsai model family, we were able to integrate support by adding the new dequantization routine without modifying the surrounding matrix operation implementations.

While our current implementations demonstrate competitive performance, they also reveal opportunities for further optimization. In particular, tighter integration between quantization layouts and GPU execution strategies, e.g., exploiting subgroup operations or specializing for specific bit-width patterns, may yield additional gains. These observations motivate future work on deeper co-design between quantization schemes and hardware-aware kernel generation.

\begin{table}[t]
\footnotesize
\caption{
Models and weight formats used for cross-device evaluation.
\texttt{llama} is additionally used in the cross-framework,
browser-vs.-native, and cross-quantization studies; the four
variants in its rows span the cross-quantization study.
}
\label{tab:models}
\begin{minipage}{\columnwidth}
\centering
\begin{tabular}{l l l l r}
\toprule
\textbf{Model} & \textbf{Short Name} & \textbf{Type} & \textbf{Weights} & \textbf{Filesize (GB)} \\
\midrule
LFM2.5-350M~\cite{amini2025lfm2technicalreport} & \texttt{lfm} & H & \texttt{q4\_k\_m} & 0.23 \\
Bonsai-1.7B~\cite{bonsai2025} & \texttt{bonsai} & B & \texttt{q1\_0} & 0.25 \\
Gemma3-270M~\cite{gemmateam2025gemma3technicalreport} & \texttt{gemma3} & T & \texttt{q4\_k\_m} & 0.25 \\
Qwen3-0.6B~\cite{yang2025qwen3technicalreport} & \texttt{qwen3} & T & \texttt{q4\_k\_m} & 0.40 \\
Granite4-1B~\cite{granite4modelcard2026} & \texttt{granite} & H & \texttt{q4\_k\_m} & 0.90 \\
Qwen3.5-2B~\cite{qwen35modelcard2026} & \texttt{qwen3.5} & T & \texttt{q4\_k\_m} & 1.28 \\
SmolLM3-3B~\cite{bakouch2025smollm3} & \texttt{smollm} & T & \texttt{q4\_k\_m} & 1.92 \\
Ministral3-3B~\cite{liu2026ministral3} & \texttt{ministral} & T & \texttt{q4\_k\_m} & 2.15 \\
Gemma4-E2B~\cite{gemma4modelcard2026} & \texttt{gemma4} & T & \texttt{q4\_k\_m} & 3.11 \\
Llama3.2-1B~\cite{llama32modelcard2024} & \texttt{llama} & T & \texttt{q2\_k}, \texttt{q4\_k\_m} & 0.58, 0.81 \\
& & & \texttt{q8\_0}, \texttt{f16} & 1.32, 2.48 \\
\bottomrule
\end{tabular}

T: Transformer.
H: Hybrid SSM.
B: 1-bit transformer.

\end{minipage}
\end{table}

\begin{table}[t]
\footnotesize
\caption{Devices used in our evaluation. Chrome is the browser used except on iOS, where Safari is the only WebGPU option.}
\label{tab:devices}
\centering

\begin{minipage}{\columnwidth}
\centering
\begin{tabular}{l r l l}
\toprule
\textbf{Vendor} & \textbf{GPUs} & \textbf{Operating Systems} & \textbf{Example Hardware} \\
\midrule
NVIDIA            & 3 & Linux, Windows   & RTX 5080$^{*\dagger}$, RTX 5070 \\
AMD               & 1 & Linux            & RX 7900 XT$^{\dagger}$ \\
Intel             & 2 & Linux, Windows   & Arc B580$^{\dagger}$, Iris Xe \\
Apple (macOS)     & 3 & macOS            & M4 (32 GB)$^{*\dagger}$, M2 \\
Apple (iOS)       & 2 & iOS              & iPhone 17 Pro Max \\
Qualcomm          & 2 & Android, Windows & Snapdragon X Elite \\
Samsung           & 1 & Android          & Galaxy S24 (Xclipse) \\
ARM               & 1 & Linux            & Mali (Valhall) \\
Imagination Tech. & 1 & Android          & PowerVR D-series \\
\midrule
\textbf{Total GPUs} & \textbf{16} & & \\
\bottomrule
\end{tabular}

\vspace{0.3em}
\raggedright
$^{*}$ Used in cross-framework memory efficiency eval. \\
$^{\dagger}$ Used in cross-framework and native performance evals. \\
\end{minipage}

\end{table}

\section{Experimental Setup}
\label{sec:eval-setup}

We evaluate LlamaWeb across a diverse set of models, model weight formats, hardware platforms, and execution environments. \mytablong~\ref{tab:models} summarizes the models and weight formats used in our study, while \mytab~\ref{tab:devices} summarizes the evaluated devices. Our evaluation spans 10 models across 16 devices from 8 GPU vendors, including desktop and mobile hardware. The evaluated models include traditional transformer architectures, recent hybrid state-space models, and a 1-bit transformer variant, allowing us to study the behavior of WebGPU inference across diverse model architectures.

Many llama.cpp quantization strategies do not uniformly quantize model weights across all layers, instead heuristically choosing a layer's weight format. Therefore, llama.cpp model variants are often named after the quantization format that is predominantly used. In particular, several strategies for \texttt{q4\_k} quantization exist. To standardize our experiments, we specifically use \texttt{q4\_k\_m} model variants for all llama.cpp \texttt{q4\_k} evaluation.

To support large-scale evaluation in the browser, we develop a benchmarking harness for the browser that automates model loading, warmup runs, repeated execution, and collection of throughput statistics. Not all models run on all devices; for example, on iOS devices Safari tab memory is limited to <500 MB. The full evaluation sweep takes anywhere from $\sim$10 minutes on a high-end discrete GPU to over an hour on low-power mobile GPUs. Across all 16 devices, cumulative time for the 481 benchmark runs in our dataset was approximately 25 hours.


For performance measurements, we separately evaluate both prefill and decode phases of inference. Prefill measurements use an input prompt length of 512 tokens, while decode measurements generate 128 output tokens. To study the impact of attention and KV-cache growth on performance, for some evaluations we measure each workload at KV-cache depths of 0 and 2048 tokens. Each experiment is repeated 5 times and we report averaged results.

All browser-based experiments are executed using WebGPU-enabled browsers on each platform. In our pilot experiments, we observed substantial implementation-dependent performance differences across browsers. Chrome and its underlying Dawn WebGPU implementation consistently provided the highest and most stable WebGPU performance across many platforms, and is therefore used for all reported  experiments unless otherwise noted. Safari is used for iOS devices where alternative browser engines are unavailable. Although Firefox supports WebGPU, we observe substantially lower performance, e.g., on an Apple M3 the \texttt{llama} model with \texttt{q4\_k\_m} weights runs at $\sim$1 tok/s on Firefox, but $\sim$52 tok/s on Chrome, so we therefore omit Firefox results from our evaluation.

\section{Memory Efficiency}
\label{sec:eval-mem}

\begin{figure}
    \centering
    \includegraphics[width=1\linewidth]{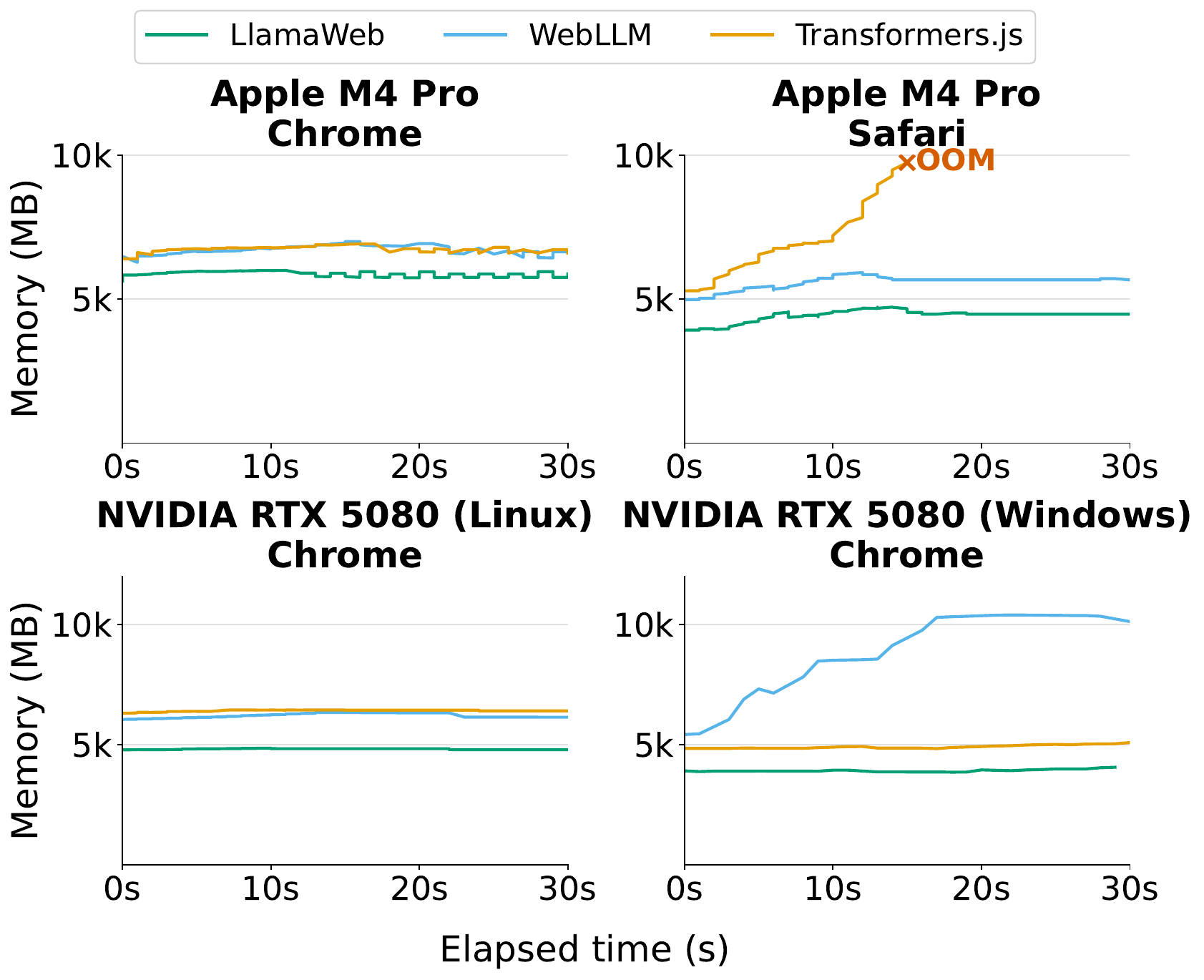}
    \caption{Memory usage for the \texttt{llama} model with \texttt{f16} weights during inference.}
    \label{fig:mem-efficiency}
\end{figure}

\myfiglong~\ref{fig:mem-efficiency} shows memory usage while running inference in the browser on the \texttt{llama} model with \texttt{f16} weights across different frameworks, browsers, and devices. We collected data using LlamaWeb, the WebLLM example webpage~\cite{webllm_chat_demo}, and Transformers.js examples~\cite{transformersjs_examples}\footnote{At evaluation time these examples had not been updated to Transformers.js 4.0, so we manually updated them to ensure we were comparing against the most recent release.}. Memory usage data was collected by launching a fresh browser instance and using platform-native tools to collect memory usage for the unified tab process for Safari, and the tab process and GPU renderer process for Chrome. Each framework was prompted to generate sufficient decode output for measurement.

In all four cases, LlamaWeb uses the least memory, with a geometric mean of normalized peak memory usage 49\% lower than WebLLM and 41\% lower than Transformers.js across the four testing configurations. On the Apple M4 Pro GPU, the memory usage of LlamaWeb was 13\% better than WebLLM and 16\% better than Transformers.js on Chrome. On Safari, this gap grew to 28\% better than WebLLM and 59\% better than Transformers.js. As the graph shows, the memory usage of Transformers.js on Safari climbed to 10 GB until the tab was eventually killed, meaning that some combination of the Transformers.js codebase and Safari's WebGPU implementation caused a memory leak. 

LlamaWeb also used less memory than other frameworks on the NVIDIA RTX 5080. Running Chrome on Linux, LlamaWeb used 24\% less memory than Transformers.js and 23\% less than WebLLM. Running Chrome on Windows, LlamaWeb used 20\% less memory than Transformers.js and 61\% less than WebLLM. WebLLM memory usage also climbed to around 10 GB on this configuration, most likely due to a memory leak. While total memory usage varies widely based on operating system, browser, and WebGPU backend, LlamaWeb consistently uses the least memory and does not suffer from memory leaks. Together, these results validate the memory-efficient design of LlamaWeb and show that developers must pay close attention to cross-platform memory usage while building applications using WebGPU.

The inefficiency of Transformers.js's steady-state memory usage can be explained in part by how it creates temporary copies of models into CPU buffers before sending them to GPU buffers, creating an unnecessarily large memory footprint. Similarly, WebLLM temporarily loads the entire model into JavaScript memory before sending it to the GPU, whereas our integration with wllama loads models into GPU buffers directly from OPFS through a small set of transfer buffers.

A last point of comparison is model sizes; llama.cpp's GGUF format and MLC's model files are often similarly sized, while ONNX models are larger due to serialization overhead. WebLLM and ONNX model files are also mandatorily sharded, by design or due to Protobuf's 2 GiB maximum file size~\cite{protobuf_docs} respectively. In contrast, GGUF models can be stored as either a single file or sharded across multiple files, giving developers more flexibility in deployment strategies.


\section{Performance Evaluations}
\label{sec:eval-perf-port}

We evaluate cross-device performance across the 10 models in our study on 16 devices, and evaluate native performance and compare performance of several WebGPU inference frameworks on an NVIDIA RTX 5080, Apple M4 Pro, AMD RX 7900 XT, and Intel Arc B580.

\subsection{Performance-Portability Across Models and Devices}
\label{sec:eval-port-models}

  \begin{figure}[t]
      \centering

      \begin{subfigure}[b]{\linewidth}
          \centering
          \includegraphics[width=\linewidth]{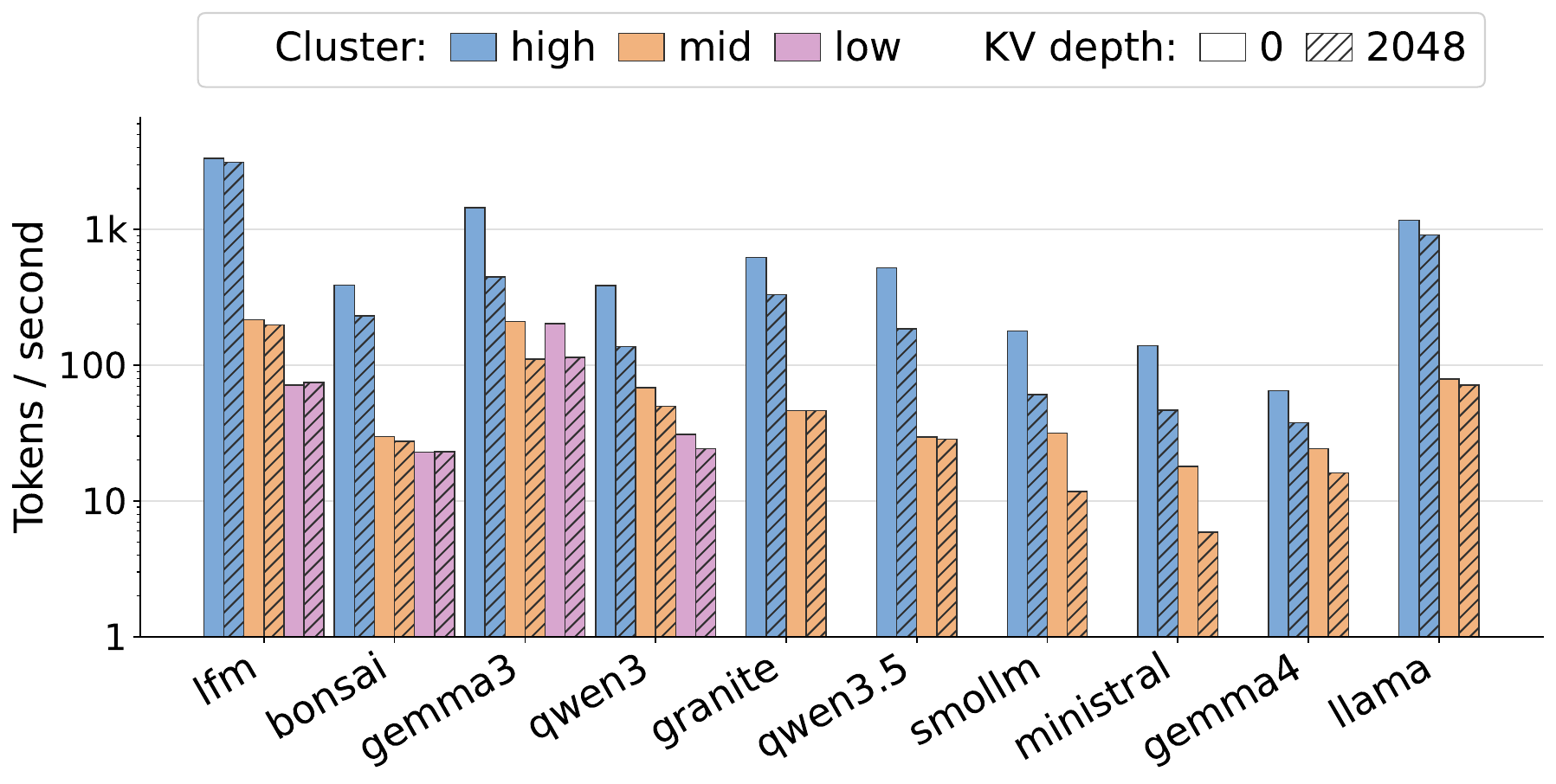}
          \caption{Prefill throughput (512-token prompt).}
          \label{fig:port-main-prefill}
      \end{subfigure}

      \vspace{0.5em}

      \begin{subfigure}[b]{\linewidth}
          \centering
          \includegraphics[width=\linewidth]{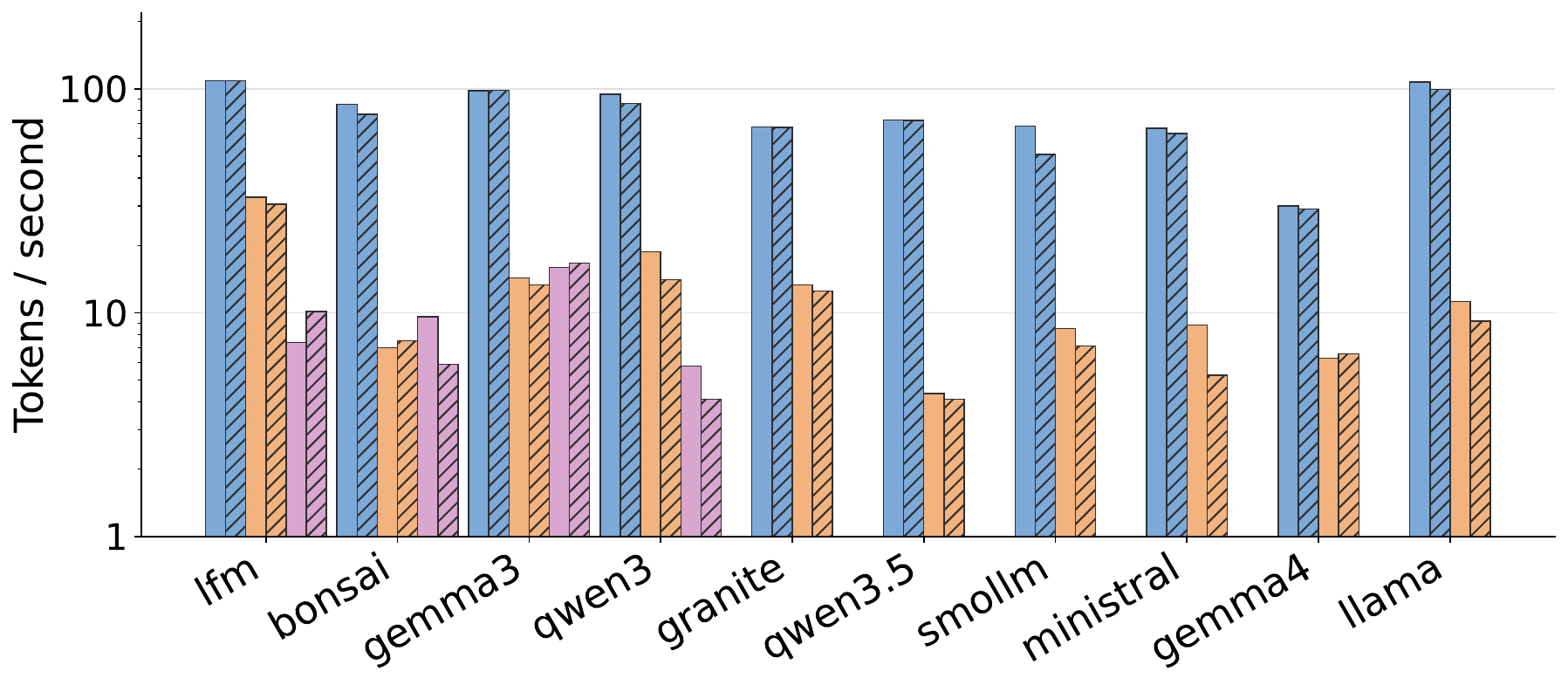}
          \caption{Decode throughput (128 generated tokens).}
          \label{fig:port-main-decode}
      \end{subfigure}
      \caption{Median LlamaWeb throughput across all 10 models (\mytab\ref{tab:models}) on 16 GPUs from 8 vendors, grouped into three
  performance clusters via $k$-means.}
      \label{fig:port-main}
  \end{figure}

To evaluate LlamaWeb's portability, we run all 10 models in \mytab\ref{tab:models} across every WebGPU-capable device in
\mytab\ref{tab:devices} with \texttt{q4\_k\_m} weights (or \texttt{q1\_0} for Bonsai) and depths of 0 and 2048 for the KV-cache. To understand how different GPUs behave and their similarities to one another, we group the devices into three clusters via $k$-means on each device's log-throughput feature vector across all (model, phase, KV-cache depth) measurements. For more details on the clustering strategy, see \myapp~\ref{app:k-means}.

The three clusters end up reasonably representing GPU capabilities and performance: the first cluster (\texttt{high}) contains high-end discrete GPUs like the NVIDIA RTX 5080, the second cluster (\texttt{mid}) contains integrated and high-end mobile GPUs like an Apple M2 and Galaxy S24, and the third cluster (\texttt{low}) contains low-power and efficient GPUs from iPhones and Android devices (Adreno, Mali, PowerVR). To our knowledge, this is the largest cross-device, cross-model evaluation of a single browser-based inference engine, and the first to include mobile devices. LlamaWeb runs every model in our suite on devices in the \texttt{high} and \texttt{mid} clusters, while GPUs in the \texttt{low} cluster were only able to fit the four smallest models (\texttt{lfm}, \texttt{bonsai}, \texttt{gemma3}, and \texttt{qwen3}) due to memory constraints.

\myfiglong~\ref{fig:port-main} shows the median prefill and decode tok/s throughput for each model, cluster, and KV-cache depth. Throughput varies by several orders of magnitude due to varied device performance and model size, highlighting how LlamaWeb is able to adapt to the extremely heterogeneous browser landscape.  On devices in the \texttt{high} cluster, smaller models reach above 3k tok/s during prefill and 100 tok/s during decode, while dropping to 65 tok/s during prefill and 30 tok/s during decode on the large \texttt{gemma4} model. Throughput on the \texttt{low} cluster is significantly lower, with decode in the range of 4--17 tok/s, but still shows that capable small models can run even in extremely constrained environments. Our results also show that while increased KV-cache depth has some effect on performance, LlamaWeb is able to scale to handle higher contexts while still achieving reasonable performance. For a more complete breakdown of model performance on every device in our study, see \myapp\ref{app:port-perdev}.

\begin{figure}
    \centering

    \begin{subfigure}[b]{\linewidth}
        \centering
        \includegraphics[width=\linewidth]{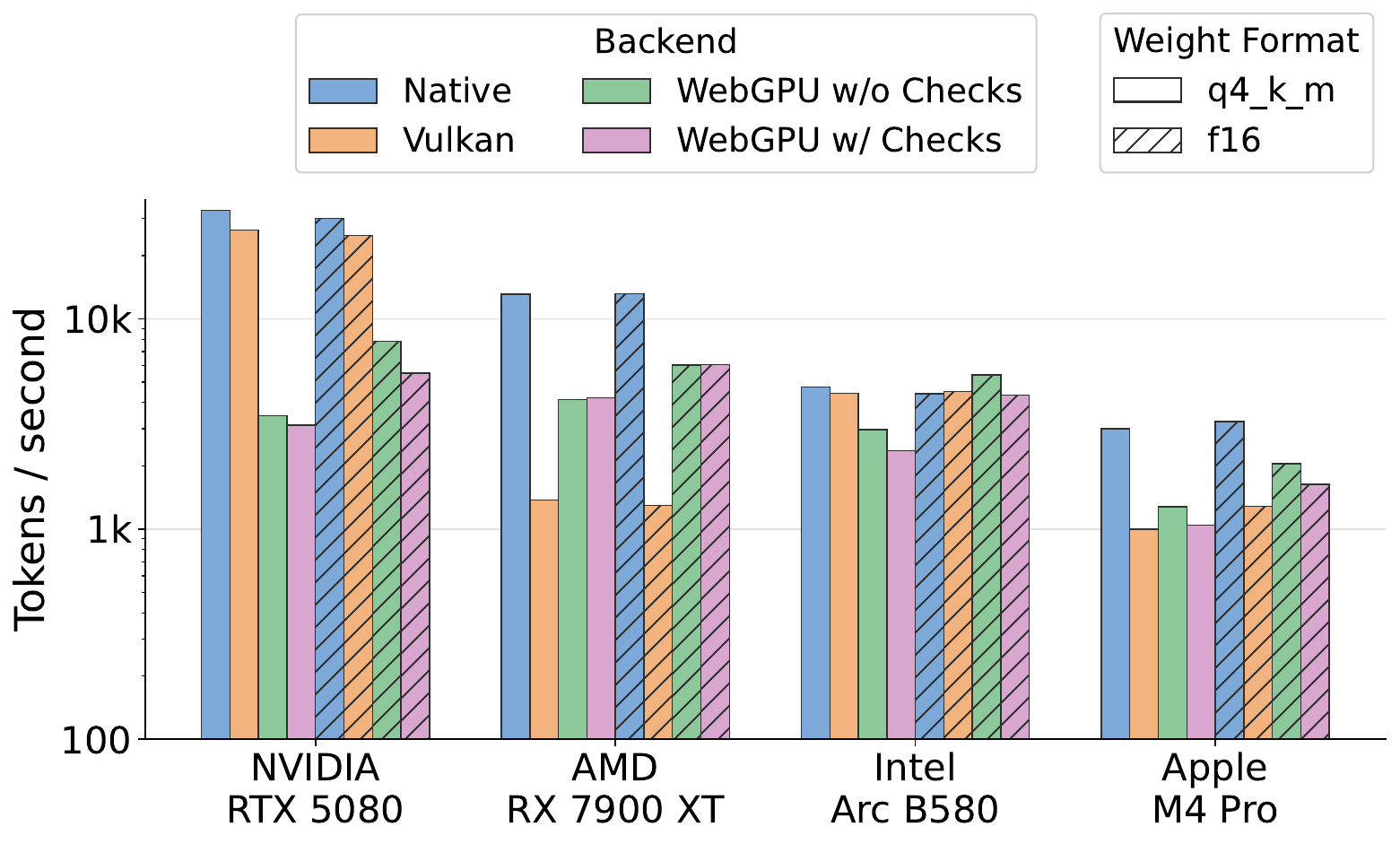}
        \caption{Prefill throughput.}
        \label{fig:native-prefill}
    \end{subfigure}

    \vspace{0.5em}

    \begin{subfigure}[b]{\linewidth}
        \centering
        \includegraphics[width=\linewidth]{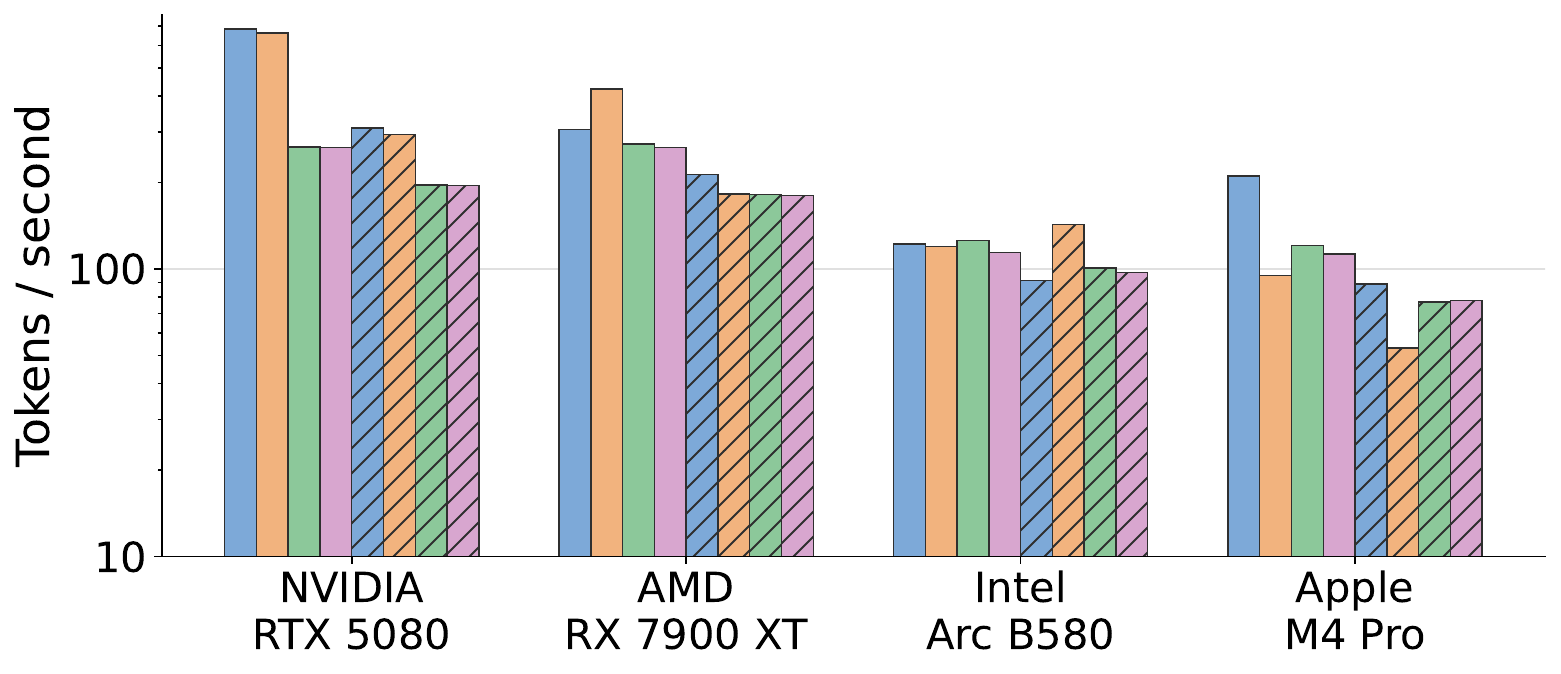}
        \caption{Decode throughput.}
        \label{fig:native-decode}
    \end{subfigure}
    \caption{Throughput of the \texttt{llama} model across different llama.cpp backends and weight formats. The native backend is CUDA on the NVIDIA GPU, HIP on the AMD GPU, SYCL on the Intel GPU, and Metal on the Apple GPU.}
    \label{fig:native}
\end{figure}

\subsection{Native Performance}
\label{sec:eval-native}

Along with the cross-device study, we also run llama.cpp natively, i.e., outside the browser, on four GPUs under several configurations as shown in ~\myfig\ref{fig:native} using llama.cpp's \texttt{llama-bench} tool. Native backends vary across devices, while Vulkan offers another cross-platform implementation for comparison against WebGPU. In the browser, WebGPU compilers add safety checks to avoid out-of-bounds buffer accesses and division-by-zero, so we also run the WebGPU backend with and without checks enabled gives a more comprehensive view of WebGPU performance. The WebGPU backend can also use the subgroup matrix feature when running natively through the Dawn WebGPU implementation, increasing the performance of our matrix multiplication and FlashAttention kernels.

As can be seen in ~\myfig\ref{fig:native-prefill}, the CUDA and Vulkan backends still significantly outperform WebGPU by up to 10$\times$ during prefill and 2.5$\times$ during decode across both model weight formats, likely due to better utilization of tensor cores and access to vendor-specific advanced features like asynchronous memory loads. Similarly, the Metal backend always outperforms the WebGPU backend, by over 2$\times$ during prefill and 50\% during decode, although the difference is not as dramatic due to the lack of tensor cores and the lower overall performance on the smaller Apple GPU. We also run Vulkan on the Apple GPU using MoltenVK~\cite{moltenvk} and find that the WebGPU backend is up to 59\% faster during prefill and 45\% faster during decode. On the AMD GPU, the results vary; while the native HIP backend does always outperform the WebGPU backend, the Vulkan backend underperforms the WebGPU backend during prefill by 3$\times$, but outperforms even the native HIP backend during decode by 38\% on the \texttt{q4\_k\_m} model. On the Intel GPU, the WebGPU backend, with safety checks disabled, provides the most performant prefill on the \texttt{f16} model, beating the SYCL backend by 23\%, and still outperforms the native SYCL backend by 10\% during decode on the same model.

Overall, these results highlight that LlamaWeb is a competitive cross-platform implementation, matching and exceeding even native backend performance in some configurations, with room to improve in others. Importantly, it shows that cross-device tuning is an important part of building a portable inference engine, as LlamaWeb's tuned matrix multiplication parameters allow it to outperform the Vulkan-specific backend on some devices, despite the fact that Vulkan is the underlying runtime for WebGPU on Linux systems. Comparing performance with and without safety checks reveals that they cause an average 14\% and 23\% slowdown during prefill on the \texttt{q4\_k\_m} and \texttt{f16} models respectively, and a 5\% and 1\% slowdown during decode. These slowdowns, which reached up to 42\% during prefill on the NVIDIA RTX 5080, show that there are opportunities for safe WebGPU to improve its performance, e.g., through better static analysis of memory accesses to remove runtime bounds-checks.

\begin{figure}
    \centering

    \begin{subfigure}[b]{\linewidth}
        \centering
        \includegraphics[width=\linewidth]{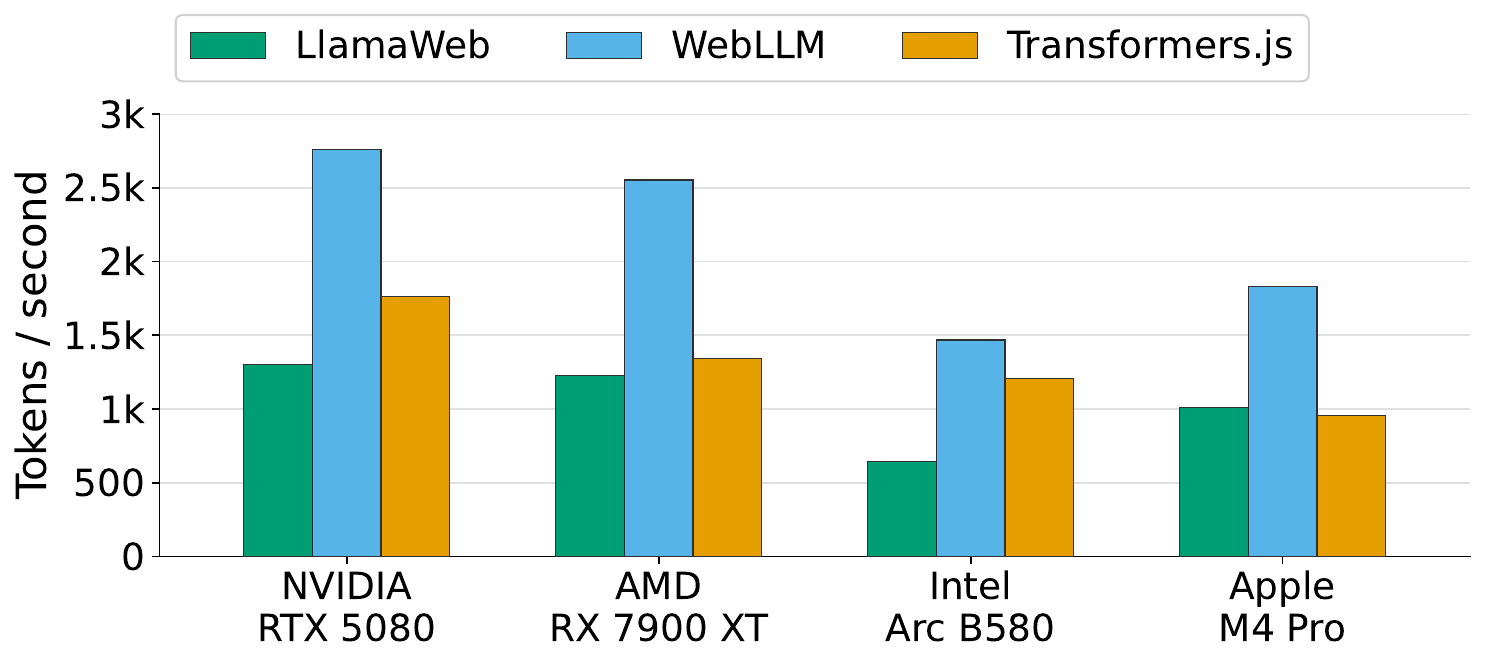}
        \caption{Prefill throughput.}
        \label{fig:perf-comp-prefill}
    \end{subfigure}

    \vspace{0.5em}

    \begin{subfigure}[b]{\linewidth}
        \centering
        \includegraphics[width=\linewidth]{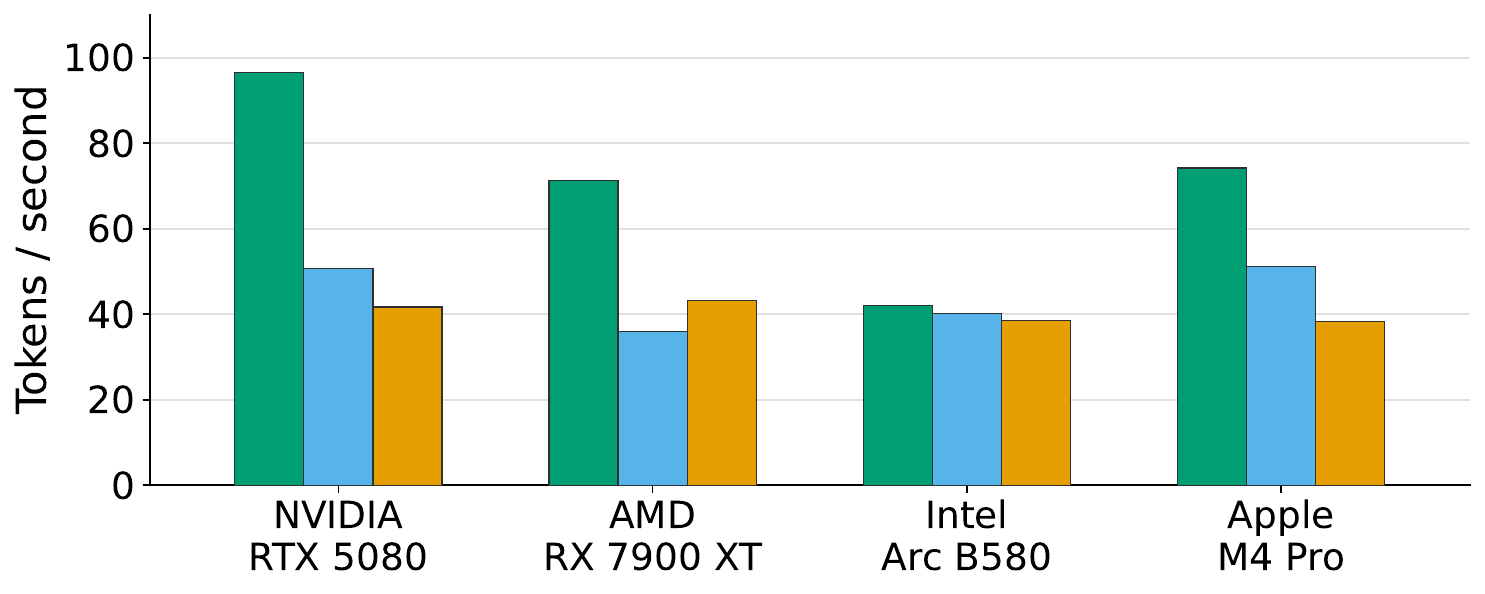}
        \caption{Decode throughput.}
        \label{fig:perf-comp-decode}
    \end{subfigure}
    \caption{Throughput comparison for the \texttt{llama} model with \texttt{f16} weights across web frameworks on Chrome. The operating system is macOS for the Apple GPU, Windows for the NVIDIA GPU, and Linux for the Intel and AMD GPUs.}
    \label{fig:perf-comp}
\end{figure}

\subsection{Framework Comparison}
\label{sec:eval-comp}
We evaluated LlamaWeb performance against other frameworks by recording prefill and decode throughput on the \texttt{llama} model with \texttt{f16} weights on Chrome on the same 4 GPUs as \mysec\ref{sec:eval-native}, with results shown in \myfig\ref{fig:perf-comp}. Because we did not have fine-grained control over benchmarking other frameworks, we used a consistent large prompt to measure prefill and a prompt that led to a relatively long decode stage to collect all measurements\footnote{For the exact prompts used, see \myapp\ref{app:comp-prompts}.}.

We calculate the geometric mean of normalized throughput across devices, as this avoids biasing devices with higher absolute throughput and correctly weights ratios. Overall, LlamaWeb achieves approximately 54\% higher decode throughput than WebLLM and 69\% higher decode throughput than Transformers.js, highlighting how our optimizations combine to provide state-of-the-art browser inference performance. However, prefill performance lags behind, with LlamaWeb achieving only 49\% of WebLLM's throughput and 79\% of Transformers.js's throughput.

The prefill gap can most likely be explained by: (1) WebLLM's use of TVM's sophisticated kernel fusion techniques to bring down the costs of loading model weights repeatedly for different small operations, and (2) Transformers.js's matrix multiplication subgroup-optimized implementation. In contrast to Transformers.js, LlamaWeb currently relies on a more portable, but less efficient register tiling matrix multiplication implementation in the browser, while our more performant kernel relying on the newer subgroup matrix feature can only run natively. In fact, using the prefill numbers from running LlamaWeb natively with checks from \mysec\ref{sec:eval-native}, LlamaWeb outperforms the prefill numbers from other frameworks on every device except WebLLM on the Apple M4 Pro, with a geometric mean speedup of 88\% over WebLLM and 205\% over Transformers.js. Therefore, although subgroup matrices are not yet available in browsers, LlamaWeb is in a strong position to provide competitive prefill performance once the feature reaches general availability. Implementing kernel fusion in LlamaWeb is also an exciting area for future work.


\section{Quantization Performance}
  \label{sec:eval-quant}

  \begin{figure}[t]
      \centering
      \begin{subfigure}[b]{\linewidth}
          \centering
          \includegraphics[width=\linewidth]{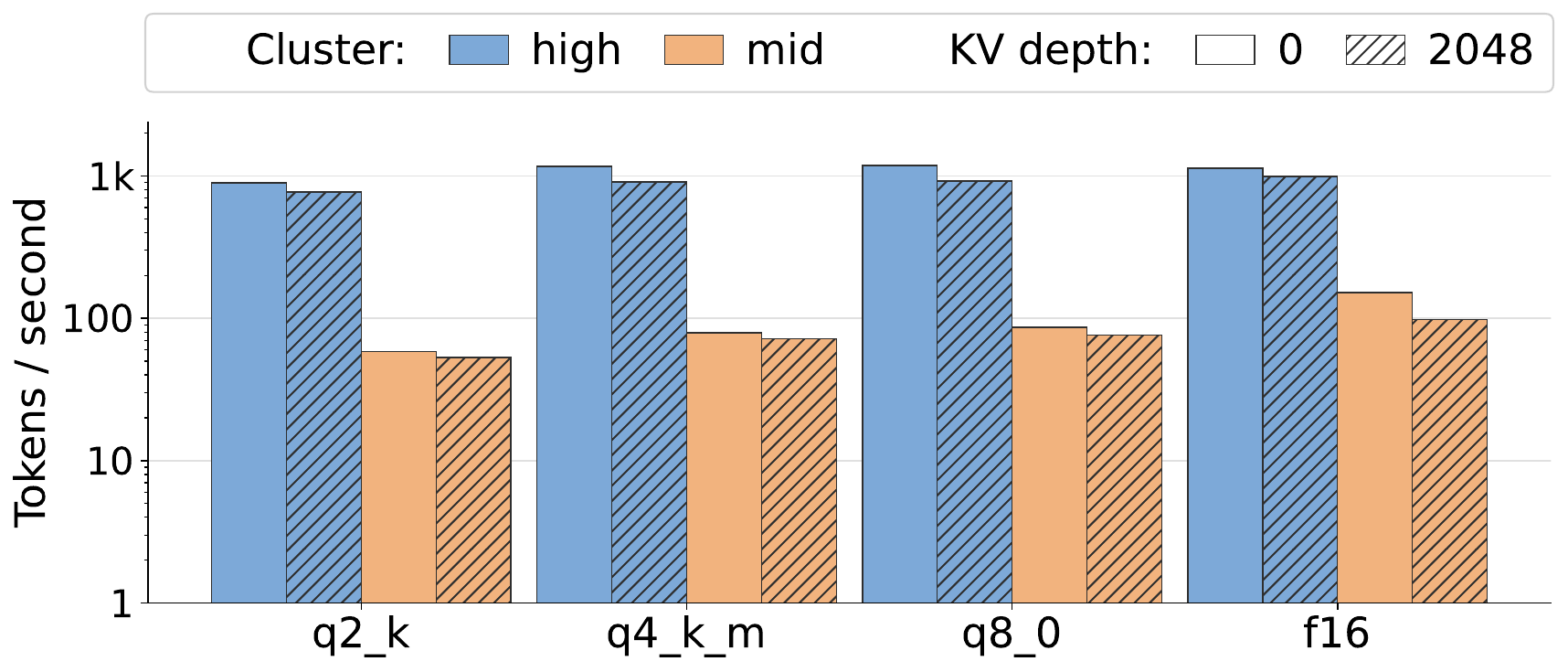}
          \caption{Prefill throughput (512-token prompt).}
          \label{fig:quant-main-prefill}
      \end{subfigure}

      \vspace{0.5em}

      \begin{subfigure}[b]{\linewidth}
          \centering
          \includegraphics[width=\linewidth]{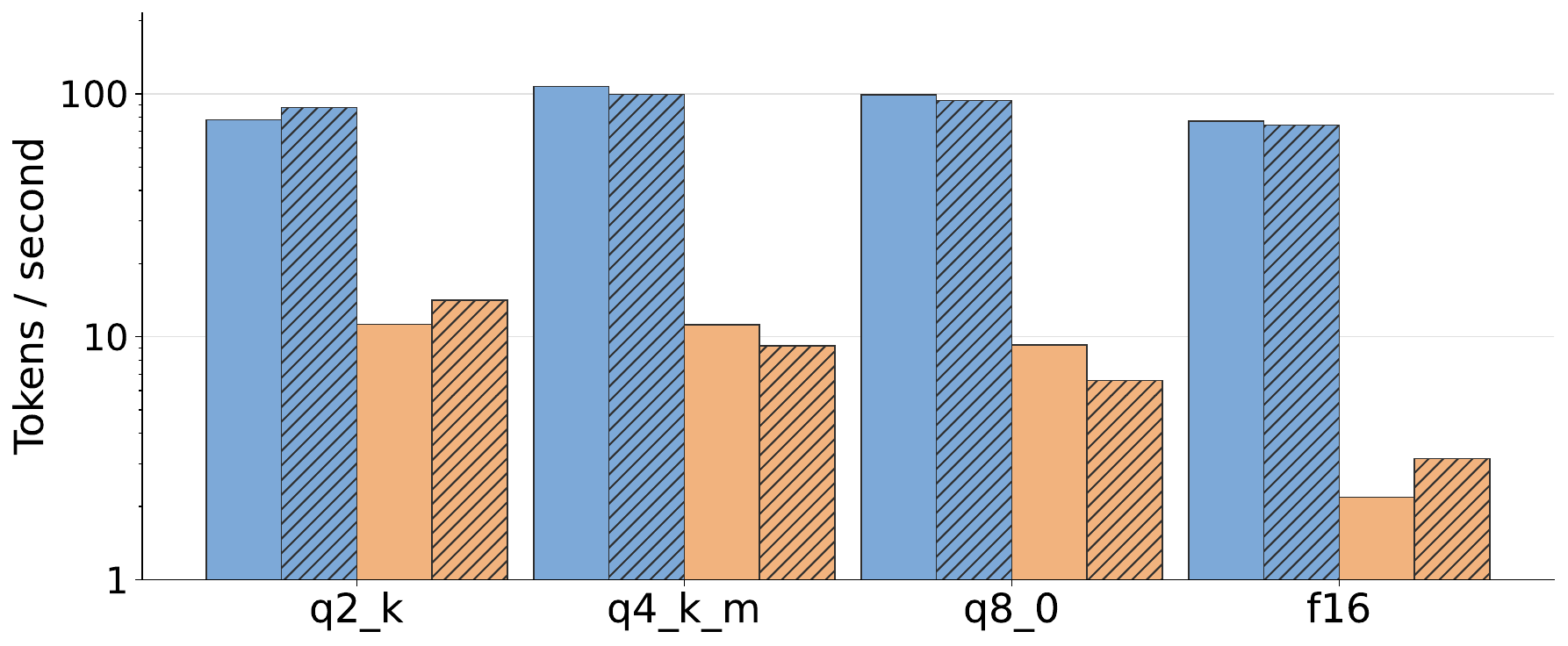}
          \caption{Decode throughput (128 generated tokens).}
          \label{fig:quant-main-decode}
      \end{subfigure}
      \caption{Throughput on the \texttt{llama} model across four weight formats (\texttt{q2\_k}, \texttt{q4\_k\_m},
  \texttt{q8\_0}, \texttt{f16}), grouped by the same device clusters as in \mysec\ref{sec:eval-port-models}.}
      \label{fig:quant-main}
  \end{figure}

To characterize how our templated dequantization kernels scale across bit widths, we run the \texttt{llama} model on the \texttt{high} and \texttt{mid} clusters from \mysec\ref{sec:eval-port-models} across four model weight formats: \texttt{q2\_k}, \texttt{q4\_k\_m}, \texttt{q8\_0}, and \texttt{f16} (\myfig\ref{fig:quant-main}). The \texttt{low} cluster is excluded from this study because the larger formats exceed Safari's iOS tab-memory budget. 

As a memory-bound operation, decode throughput (\myfig\ref{fig:quant-main-decode}) shows a clear speedup when moving from \texttt{f16} to \texttt{q8\_0} weights, e.g., 20\% on the \texttt{high} cluster and 53\% on the \texttt{mid} cluster with a KV-cache depth of 0. However, performance does not scale linearly when moving to lower-bit quantization formats; while the lower memory traffic does lead to an 8\% increase in throughput when moving from \texttt{q8\_0} to \texttt{q4\_k\_m} weights on \texttt{mid} cluster, moving from \texttt{q4\_k\_m} to \texttt{q2\_k} weights causes a 17\% decrease in throughput on the same cluster. Prefill (\myfig\ref{fig:quant-main-prefill}) shows a flatter profile, and the \texttt{f16} weight format actually performs best on the \texttt{mid} cluster, with a median throughput of 152 tok/s vs. 79 for \texttt{q4\_k\_m}. Generally, performance across model weight format stays within an order of magnitude, and the gap between the fastest and slowest model weight format for prefill is only 25\% on the \texttt{high} cluster.


KV-cache growth has a similar effect across formats. Prefill drops 10--20\% from depth 0 to 2048 on the \texttt{high} cluster regardless of weight
format, consistent with the attention computation becoming proportionally larger. Decode is barely affected: median \texttt{q4\_k} throughput only falls
from 107 to 99 tok/s on the \texttt{high} cluster, and the other formats track within a few percent. Once a model weight format is chosen, KV-cache scaling behavior is fairly predictable.

These results show that as it stands llama.cpp quantization in WebGPU is mainly a technique for reducing the memory requirements for running a given model, rather than a performance optimization. Dequantization routines for generic model weight formats are complex and computationally heavy compared to hardware-native formats like \texttt{nvfp4}, which WebGPU does not support, and the current layouts may not be optimized for performance on the variety of GPUs that WebGPU targets. However, the results do validate that the templated kernels introduced in this work deliver competitive performance across various formats, and lay a foundation for future optimization work. For a more complete breakdown of model weight format performance across devices, see \myapp\ref{app:quant-perdev}.


\section{Related Work}
\label{sec:related}

\paragraph{LLM Inference Engines}
In the browser, WebLLM and Transformers.js provide WebGPU-based inference through MLC-LLM and ONNX Runtime, and we compare against these frameworks in our evaluation. WeInfer~\cite{chen2025weinfer} improves aspects of WebLLM's execution scheduling, but does not target the same broader goals around memory efficiency, performance portability, and quantization support explored in this work. MediaPipe~\cite{lugaresi2019mediapipeframeworkbuildingperception} also supports WebGPU inference, though current examples and documentation focus primarily on Gemma-family models. Outside the browser, systems such as MLX~\cite{mlx2023}, vLLM~\cite{kwon2023vllm}, and TensorRT-LLM~\cite{tensorrtllm} target native GPU backends and high-throughput server inference, often emphasizing batching and datacenter deployment rather than execution on consumer and edge devices.

\paragraph{Performance Portability}
The existing browser-based inference frameworks we evaluate do specialize some kernels based on device characteristics, but their limited performance portability strategies are generally undocumented. Beyond WebGPU, prior work has formalized the notion of performance portability and proposes quantitative metrics for measuring performance gaps~\cite{pennycook2016perfport}. Portable GPU libraries such as CLBlast expose numerous tuning parameters per-kernel and maintain community-contributed databases of kernel performance~\cite{Nugteren_2018}. Auto-tuning frameworks like GPTune explore large optimization spaces for parallel applications and leverage performance data collected from many machines to guide future tuning decisions~\cite{liu2021gptune}. These techniques are complementary to the design goals of the LlamaWeb kernel library, as it can evolve to support more sophisticated tuning strategies and integration with kernel tuning frameworks.

\paragraph{Quantization}
Quantization approaches can be broken down into training and post-training methods. In quantization-aware training, weights are learned while taking into account quantization effects~\cite{jacob2017qat, esser2020learnedstepsizequantization, courbariaux2015trainingdeepneuralnetworks}. Post-training quantization (PTQ) includes weight-only methods such as GPTQ~\cite{frantar2023gptq}, which does substantial offline analysis and optimization while modifying weights, and AWQ~\cite{lin2026awq}, which does a more lightweight statistics-based analysis. Runtime-aware PTQ methods like SmoothQuant~\cite{xiao2024smoothquant} and LLM.int8()~\cite{dettmers2022llmint8} combine offline analysis with runtime computation like activation scaling or mixed-precision execution. Quantization work is often based around developing new quantization strategies, rather than integrating them into a single runtime and templated kernels like this work. Support for various model weight formats and developing extensible kernel formats like we do in this work is an important step towards the development and deployment of novel quantization techniques.

\section{Future Work}
\label{sec:future}

This work is only the starting point for LlamaWeb and inference in the browser using WebGPU. Within LlamaWeb itself, interesting future work opportunities include: (1) advances in performance-portable tuning techniques to enable maximal WebGPU performance on diverse systems, (2) new techniques for handling the functional and performance requirements of diverse quantization formats, (3) dynamic kernel fusion to reduce the overhead of repeatedly loading memory and launching many small GPU kernels, and (4) support for mixture-of-expert and multi-modal models in WebGPU (with several llama.cpp contributors already making strides in this direction).

There are also opportunities to optimize WebGPU implementations and improve the specification itself to better support applications like LLM inference in the browser, including: (1) static analysis techniques and optimizations to avoid safety and bounds-checks, which decrease performance in the browser, (2) more robust language-level handling of floating-point differences across systems to ensure model stability under lower-precision execution, and (3) extensions to WebGPU to support new data-types, e.g., \texttt{u16}, that enable more efficient memory loading and native hardware acceleration.



\section{Conclusion}
\label{sec:conclusion}

In this work, we introduce LlamaWeb, a WebGPU backend for llama.cpp that is designed with goals of memory efficiency, performance portability, and broad model weight format support in mind. We show that compared to existing frameworks, LlamaWeb provides state-of-the-art results with respect to these goals. LlamaWeb provides a new opportunity for accessible and performant browser-based LLM inference, with several outside contributors already contributing to and utilizing it for exciting new use-cases. We look forward to seeing how the combined llama.cpp and WebGPU ecosystems continue to evolve based upon the foundation laid in this work.

\section{Acknowledgments}

We thank the maintainers and developers of llama.cpp for their support in helping us integrate the WebGPU backend into llama.cpp. We also thank the outside contributors who are already contributing to and improving the llama.cpp WebGPU backend. This work was supported in part by an NDSEG fellowship.

\bibliographystyle{ACM-Reference-Format}
\bibliography{references}

\clearpage

\appendix

\section{Number of Available Models}
\label{app:model-support}
We searched Hugging Face for available models for each framework at the time of submission. LlamaWeb loads models in the GGUF format, of which there were 177,691 on Hugging Face in May 2026. While not all of these models are likely to run in the browser due to large sizes or legacy formats, our tests and evaluation show that every recent model we tested did run successfully on large-enough GPUs. Transformers.js uses ONNX Runtime as its backend, and there were 41,632 ONNX models available on Hugging Face. As with LlamaWeb, not all of these models are likely to be compatible with Transformers.js, but to provide a fair comparison we use this number in \mytab\ref{tab:cool-info}. WebLLM only directly supports models in the MLC format, and the Hugging Face mlc-ai model repository currently contains 400 models in this format.

\section{Framework Model Weight Support}
\label{app:quant-support}

In this section we briefly discuss the model weight formats available for inference in each of the three frameworks we evaluate in this work. Within each framework, a single model and its activations may be stored in several different formats, depending on how the model is quantized. However, we do not count each of these potential mixtures separately, and focus only on the core underlying data formats.

\paragraph{LlamaWeb} Our implementation supports \texttt{f32} and \texttt{f16} floating-point values. As discussed in \mysec~\ref{sec:background}, WebGPU does not include support for some native floating-point formats. We also support the three main quantization families in llama.cpp: legacy, including \texttt{q4\_0}, \texttt{q4\_1}, \texttt{q5\_0}, \texttt{q5\_1}, and \texttt{q8\_0}, K-quants, including \texttt{q2\_k}, \texttt{q3\_k}, \texttt{q4\_k}, \texttt{q5\_k}, and \texttt{q6\_k}, and I-quants, including \texttt{iq1\_s}, \texttt{iq1\_m}, \texttt{iq2\_xxs}, \texttt{iq2\_xs}, \texttt{iq2\_s}, \texttt{iq3\_xxs}, \texttt{iq3\_s}, \texttt{iq4\_nl}, and \texttt{iq4\_xs}. Finally, we recently added support for the new \texttt{q1\_0} format, and an outside collaborator added support for \texttt{mxfp4}. There are several other formats supported by llama.cpp that the WebGPU does not yet support, but which may be able to be added in the future, including 64-bit weights and ternary quantization formats.

\paragraph{WebLLM} By investigating the MLC-LLM codebase~\cite{mlc-llm}, which powers WebLLM and uses TVM under the hood, and specifically its \texttt{quantization.py} file, we determined that WebLLM currently supports 6 formats, including \texttt{f32}, \texttt{f16}, \texttt{bf16}, \texttt{f8}, and 3 and 4-bit formats. To support hardware-native formats like \texttt{bf16} and \texttt{f8}, which are not available in WebGPU, MLC-LLM generates WGSL code that ``legalizes'' the weights at runtime through bit-interpretation into types like \texttt{f32} and \texttt{f16} that can be executed by WebGPU. In practice, e.g., in the hosted WebLLM examples~\cite{webllm_chat_demo}, \texttt{f32}, \texttt{f16}, and 4-bit quantization seem to be the most widely deployed model weight formats.

\paragraph{Transformers.js} We investigated the Transformers.js~\cite{transformersjs} and ONNX Runtime~\cite{onnxruntime} codebases to determine their support for different weight formats. There is no centralized list of quantization format for the ONNX Runtime WebGPU backend, but we determined that it likely supports \texttt{f32}, \texttt{f16}, signed and unsigned 8 and 4-bit types interpreted directly as floating point values, as well as block or tensor-wise 4 and 2-bit quantization. In Transformers.js, \texttt{dtypes.js} lists support for \texttt{f32}, \texttt{f16}, signed and unsigned 8-bit types, 8, 4, 2, and 1-bit quantization, and a special type called \texttt{bnb4}. To our knowledge, \texttt{bnb4} and 8-bit quantization do not have paths to the ONNX Runtime WebGPU backend, while 1-bit quantization is implemented by expanding weights at runtime into a 2-bit representation that is compatible with the ONNX Runtime WebGPU kernels. This means that while 1-bit models are still stored in a compact form, they incur a memory overhead at runtime. Overall, this leads us to conclude that 7 weight formats are supported for inference in WebGPU by Transformers.js.

In practice, we find that like WebLLM, the publicly available Transformers.js examples~\cite{transformersjs_examples} tend to use \texttt{f32}, \texttt{f16}, and 4-bit quantization model weight formats.

\begin{table}[t]
\small
\caption{Cluster membership after $k$-means on log-throughput feature vectors with $k=3$, as analyzed in \mysec\ref{sec:eval-port-models}.}
\label{tab:cluster-membership}
\centering
\begin{tabular}{l p{0.66\columnwidth}}
\toprule
\textbf{Cluster} & \textbf{Devices} \\
\midrule
\texttt{high} & RTX 5080, RTX 5070, RTX 40-series, RX 7900 XT, Arc B580, M4 (32~GB), M3 (16~GB) \\
\texttt{mid}  & Iris Xe (iGPU), M2, Snapdragon X Elite, Galaxy S24 \\
\texttt{low}  & iPhone 17 Pro Max, iPhone 15, Adreno 7xx, Mali (Valhall), PowerVR D-series \\
\bottomrule
\end{tabular}
\end{table}
  
\section{Clustering GPUs with \textit{k}-Means}
\label{app:k-means}
To understand how LlamaWeb performs on GPUs across vendors and device characteristics, we group the 16 GPUs in our evaluation in \mysec\ref{sec:eval-port-models} into three clusters via $k$-means. This appendix documents how the feature vectors are constructed, how $k$ was chosen, and which devices end up in each cluster (\mytab\ref{tab:cluster-membership}).

\paragraph{Feature Vectors}
Each device contributes one row to the input matrix. Columns are the per-(model, phase, KV-depth) measurements collected during the
cross-device study: 10 models $\times$ 2 phases (prefill, decode) $\times$ 2 KV-cache depths (0 and 2048) yields 40 feature columns. Cell
values are $\log1p(\text{throughput})$ in tokens/sec; the log transform normalizes throughput values which span more than three orders of magnitude. When a model hasn't run on a device (e.g.\ \texttt{gemma4} on an iPhone with a tight tab-memory cap), we impute the missing cell with the per-column median across all devices that did run it. 


\paragraph{Choice of $k$ and Cluster Membership}
We fit $k$-means with $k \in \{2, 3, 4, 5\}$ on the 16-device feature matrix. Inertia decreases without a sharp elbow (358 $\to$
272 $\to$ 223 $\to$ 173), highlighting the inherent heterogeneity of devices supported by WebGPU. For analysis in this paper, we set $k=3$, as it leads to an intuitive split of devices as shown in \mytab\ref{tab:cluster-membership}. The \texttt{high} cluster captures the high-end discrete GPUs (RTX 5080/5070/40-series, RX 7900 XT, Arc B580) and the higher-tier Apple silicon (M3, M4). The \texttt{mid} cluster collects integrated and high-end-mobile GPUs (Iris Xe, M2, Snapdragon X Elite, Galaxy S24). The \texttt{low} cluster contains iOS devices and the low-power Android GPUs (Adreno 7xx, Mali (Valhall), PowerVR D-series, iPhone 15, iPhone 17 Pro Max).



\section{Per-Device Portability Results}
  \label{app:port-perdev}

  This appendix supplements \mysec\ref{sec:eval-port-models} with full per-device data for the portability study. \myfiglong\ref{fig:port-coverage-prefill} shows the coverage and prefill throughput across each model and device in our study, while \myfig\ref{fig:port-coverage-decode} shows the decode throughput. Dark blue cells correspond to higher throughput, light blue to lower throughput, and cells where the given model-device combo didn't run are left blank.
  
\begin{figure*}
\centering
\setlength{\tabcolsep}{2pt}
\renewcommand{\arraystretch}{0.95}

\begin{tabular}{cc}

\subfloat[\texttt{lfm} prefill]{\includegraphics[width=0.48\textwidth]{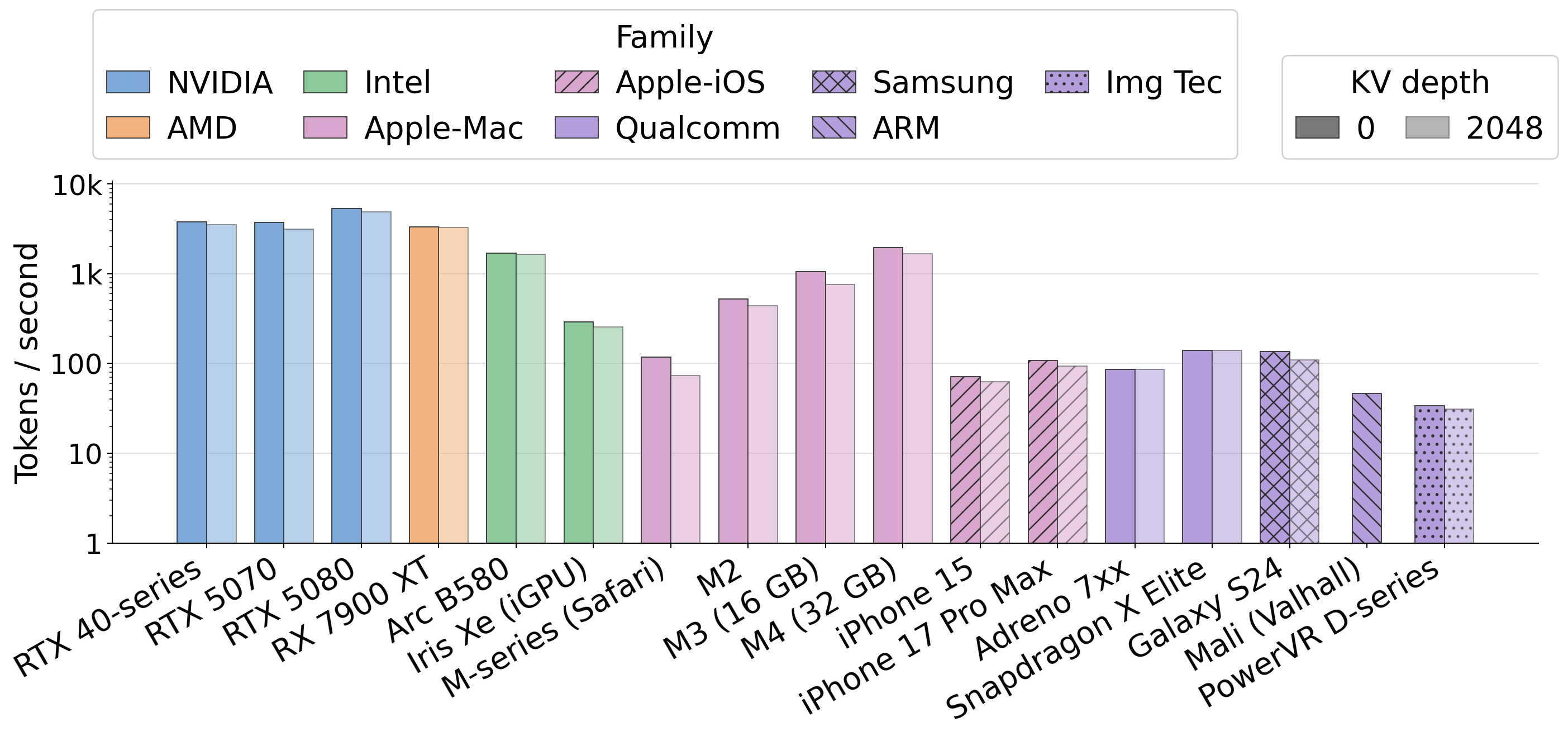}} &
\subfloat[\texttt{lfm} decode]{\includegraphics[width=0.48\textwidth]{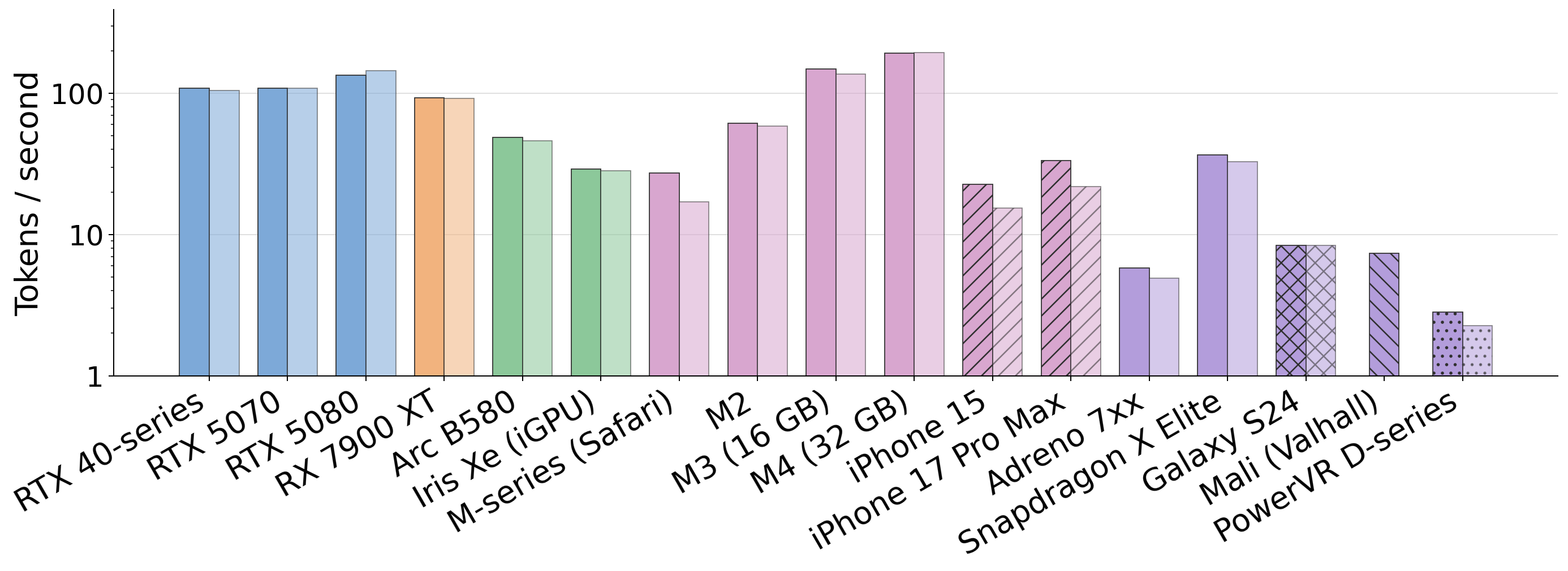}} \\

\subfloat[\texttt{bonsai} prefill]{\includegraphics[width=0.48\textwidth]{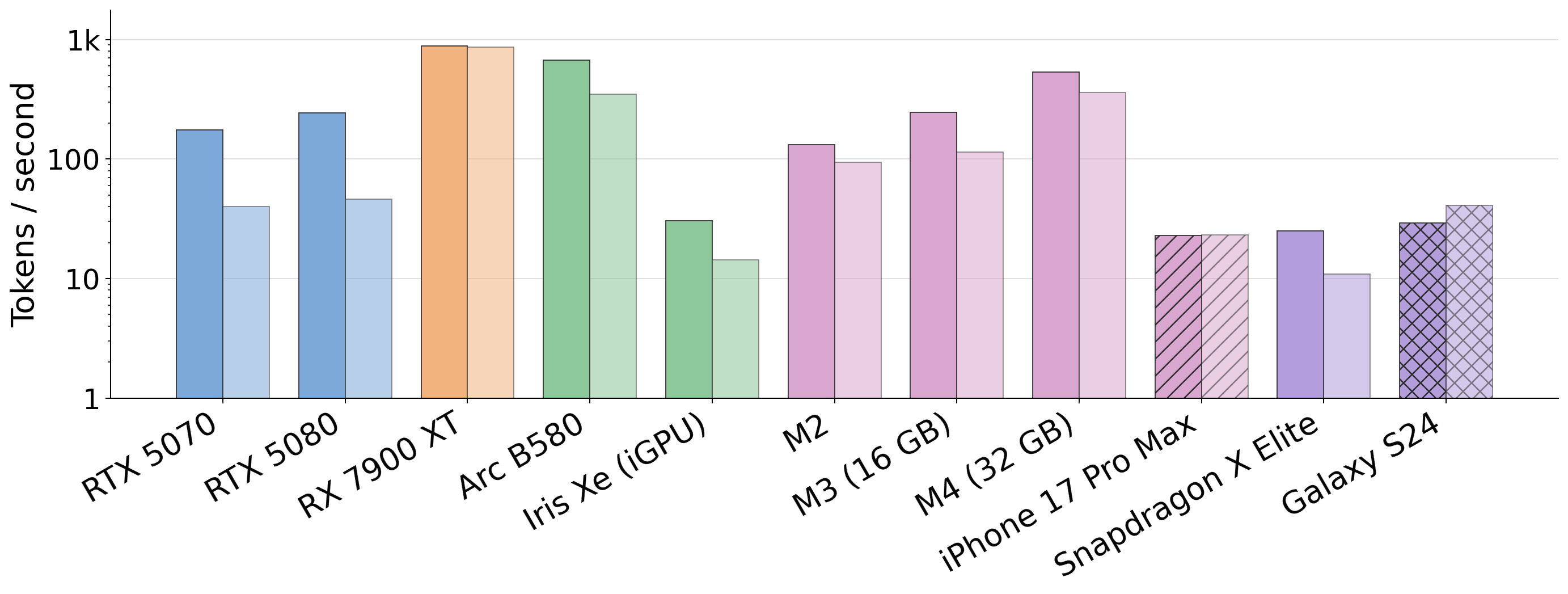}} &
\subfloat[\texttt{bonsai} decode]{\includegraphics[width=0.48\textwidth]{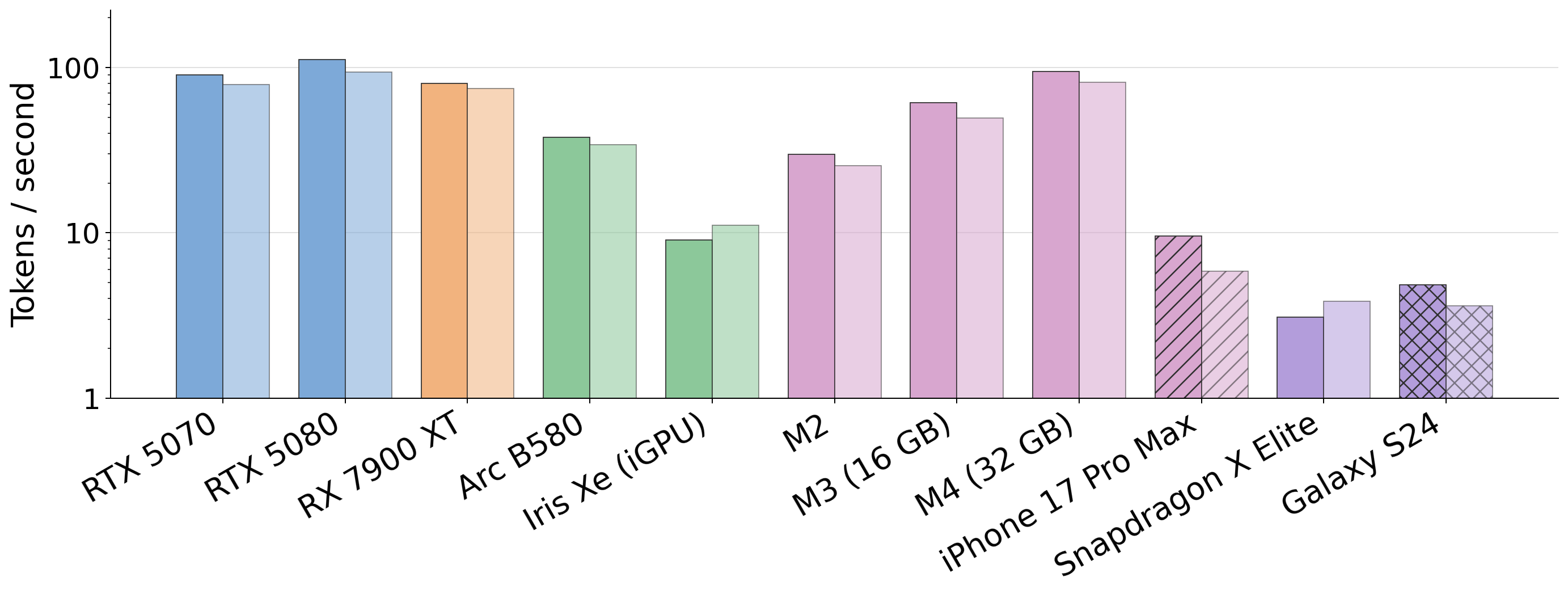}} \\

\subfloat[\texttt{gemma3} prefill]{\includegraphics[width=0.48\textwidth]{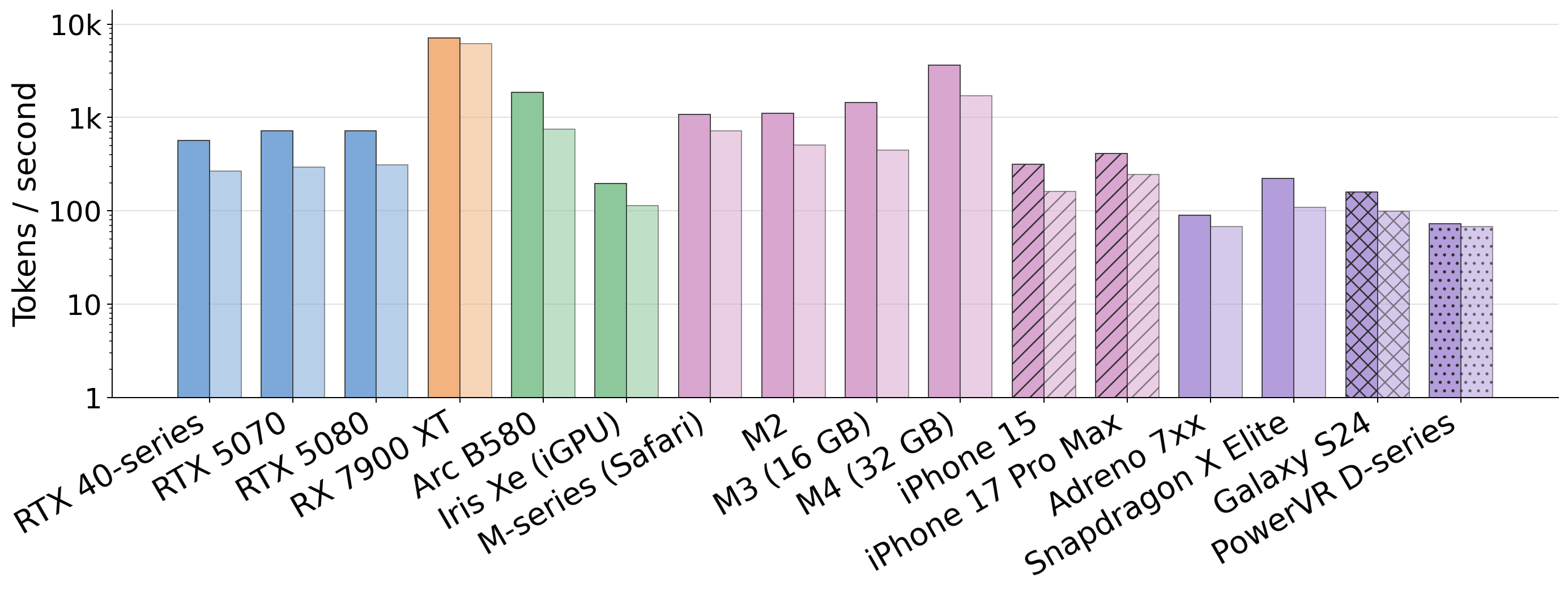}} &
\subfloat[\texttt{gemma3} decode]{\includegraphics[width=0.48\textwidth]{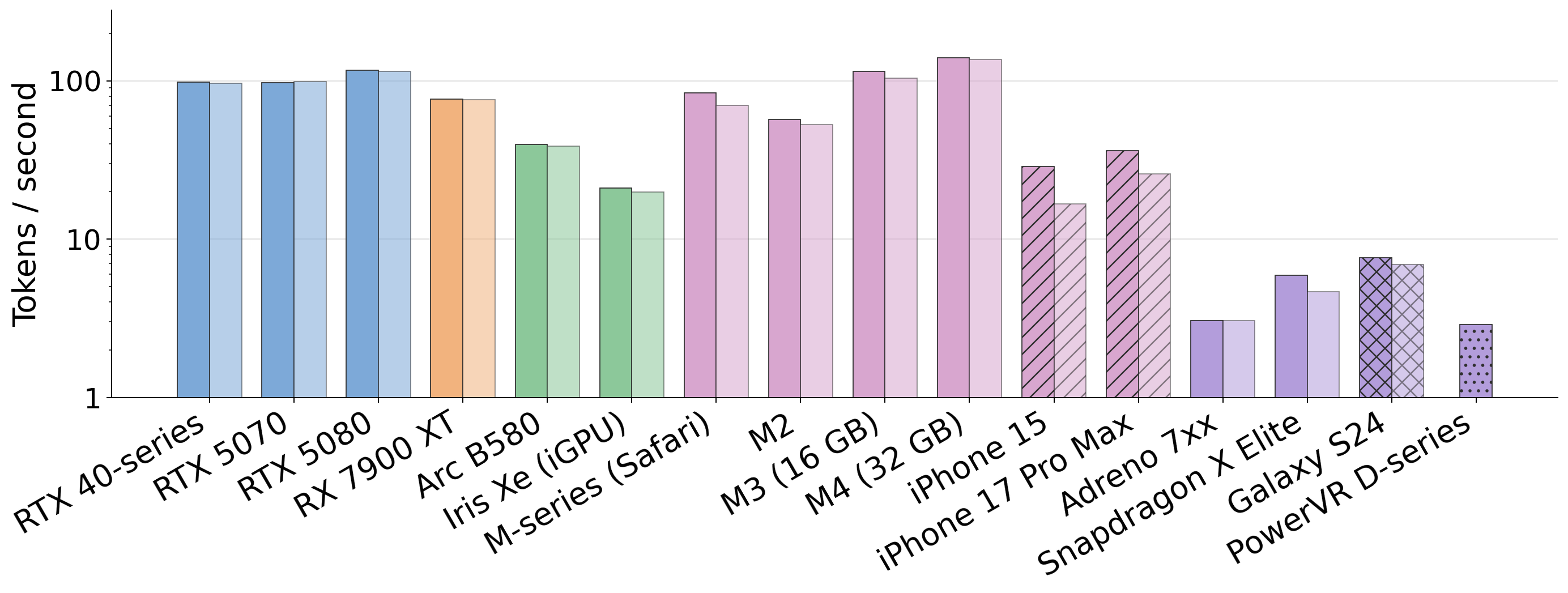}} \\

\subfloat[\texttt{qwen3} prefill]{\includegraphics[width=0.48\textwidth]{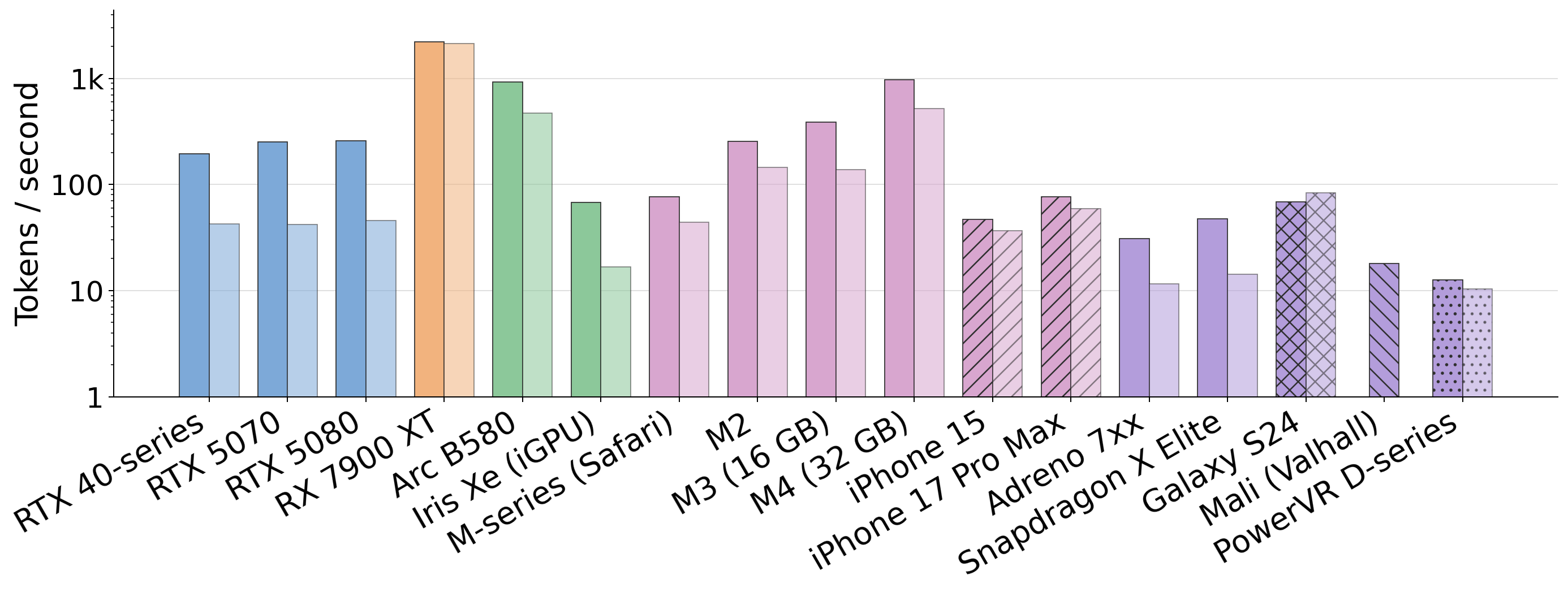}} &
\subfloat[\texttt{qwen3} decode]{\includegraphics[width=0.48\textwidth]{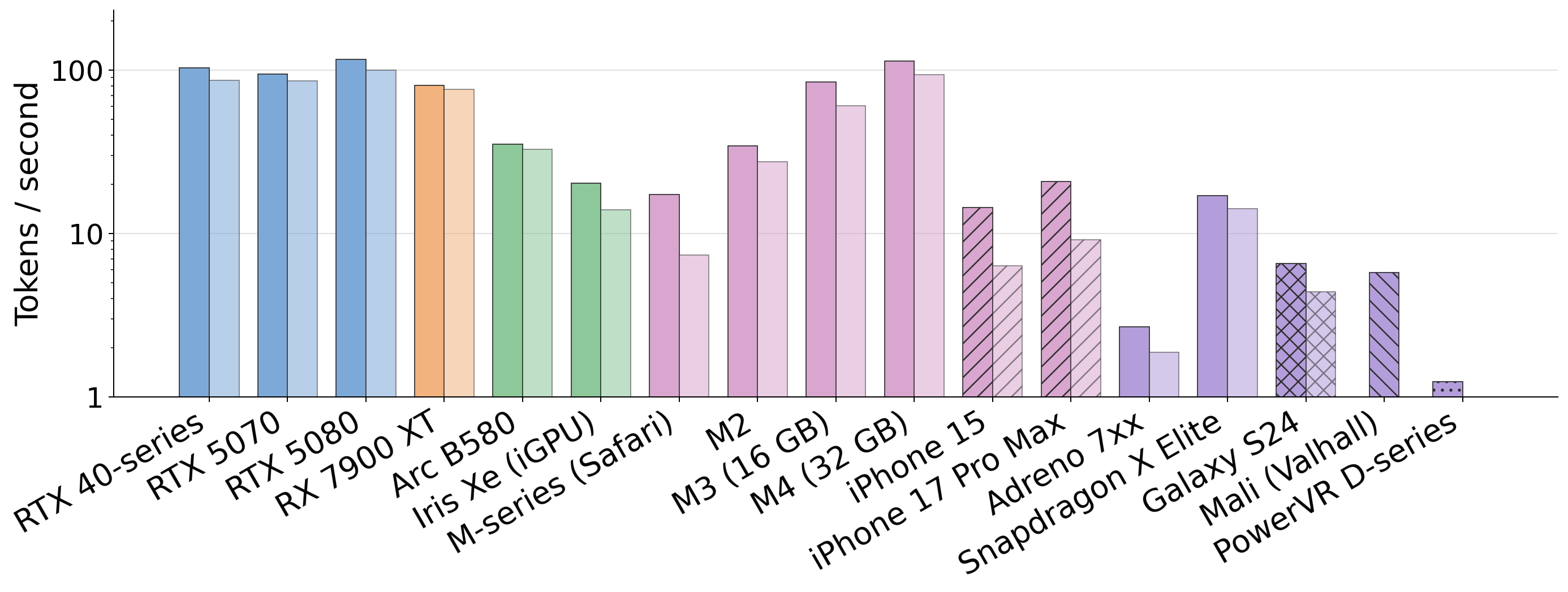}} \\

\subfloat[\texttt{granite} prefill]{\includegraphics[width=0.48\textwidth]{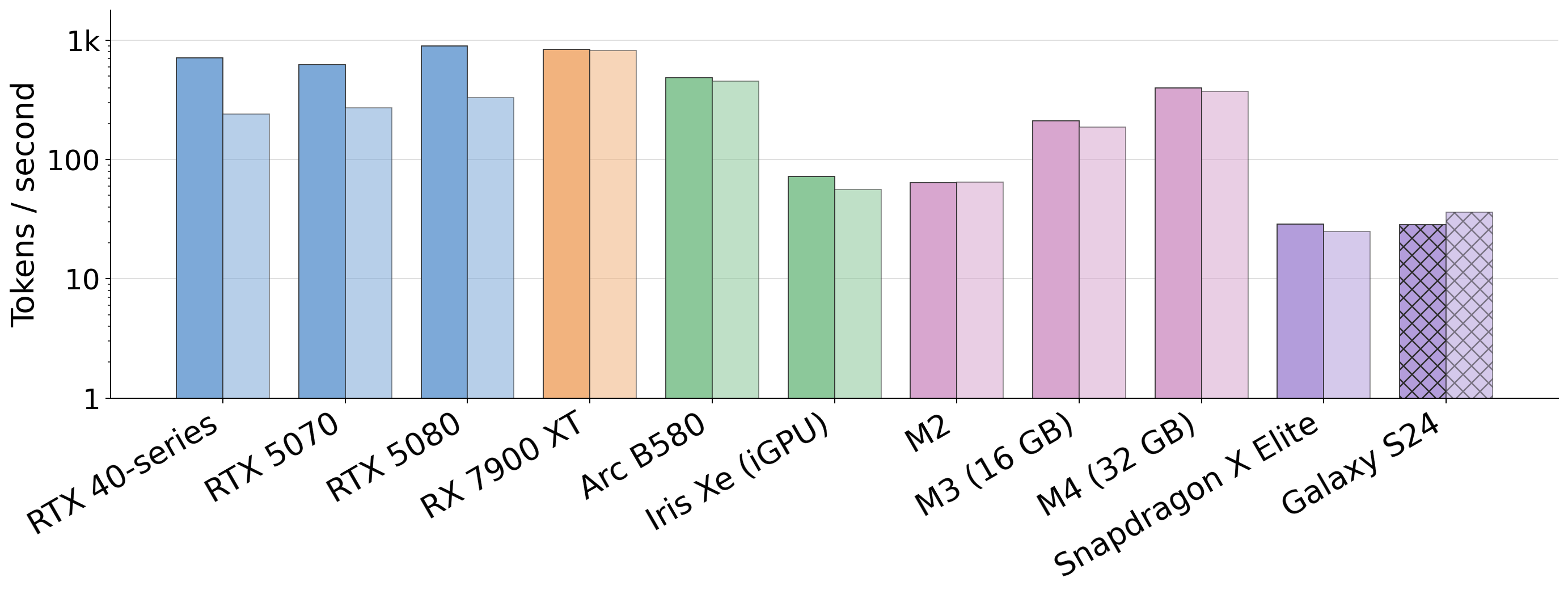}} &
\subfloat[\texttt{granite} decode]{\includegraphics[width=0.48\textwidth]{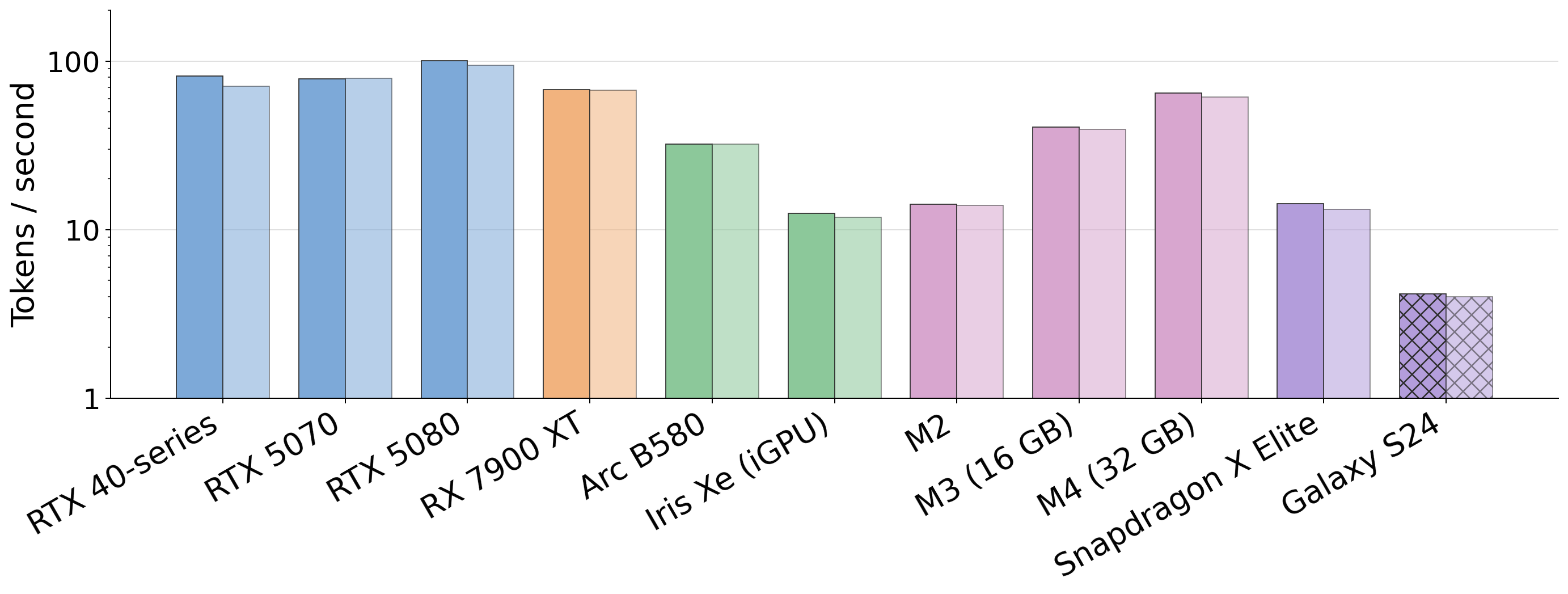}} \\

\end{tabular}

\caption{Per-device prefill and decode throughput by model, part 1 of 2.}
\label{fig:port-perdev-1}
\end{figure*}

\begin{figure*}
\centering
\setlength{\tabcolsep}{2pt}
\renewcommand{\arraystretch}{0.95}

\begin{tabular}{cc}

\subfloat[\texttt{qwen3.5} prefill]{\includegraphics[width=0.48\textwidth]{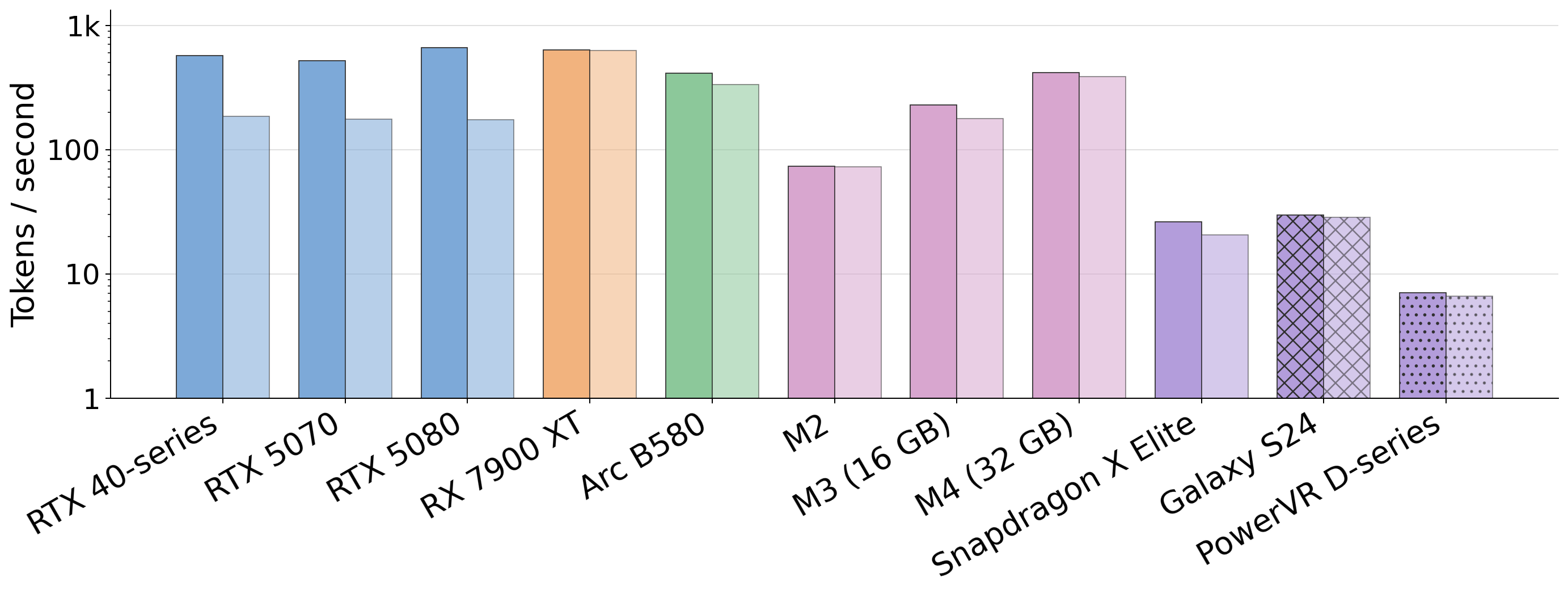}} &
\subfloat[\texttt{qwen3.5} decode]{\includegraphics[width=0.48\textwidth]{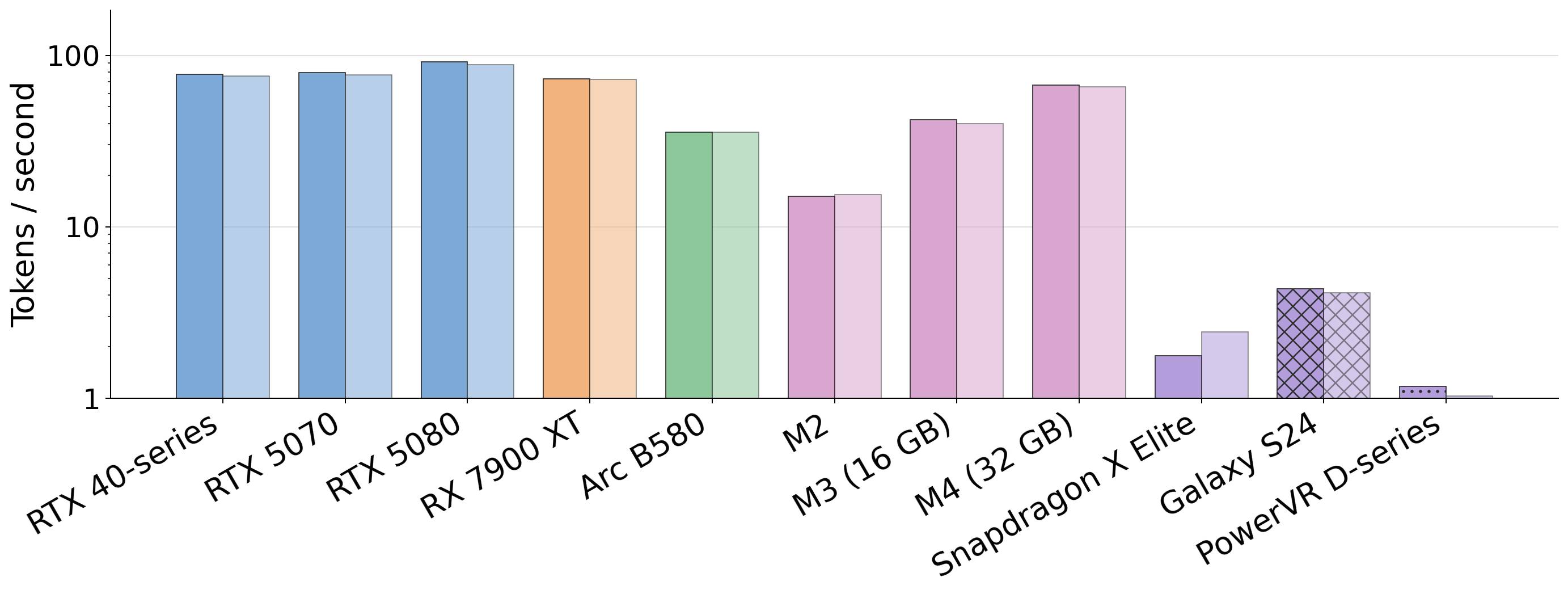}} \\

\subfloat[\texttt{smollm} prefill]{\includegraphics[width=0.48\textwidth]{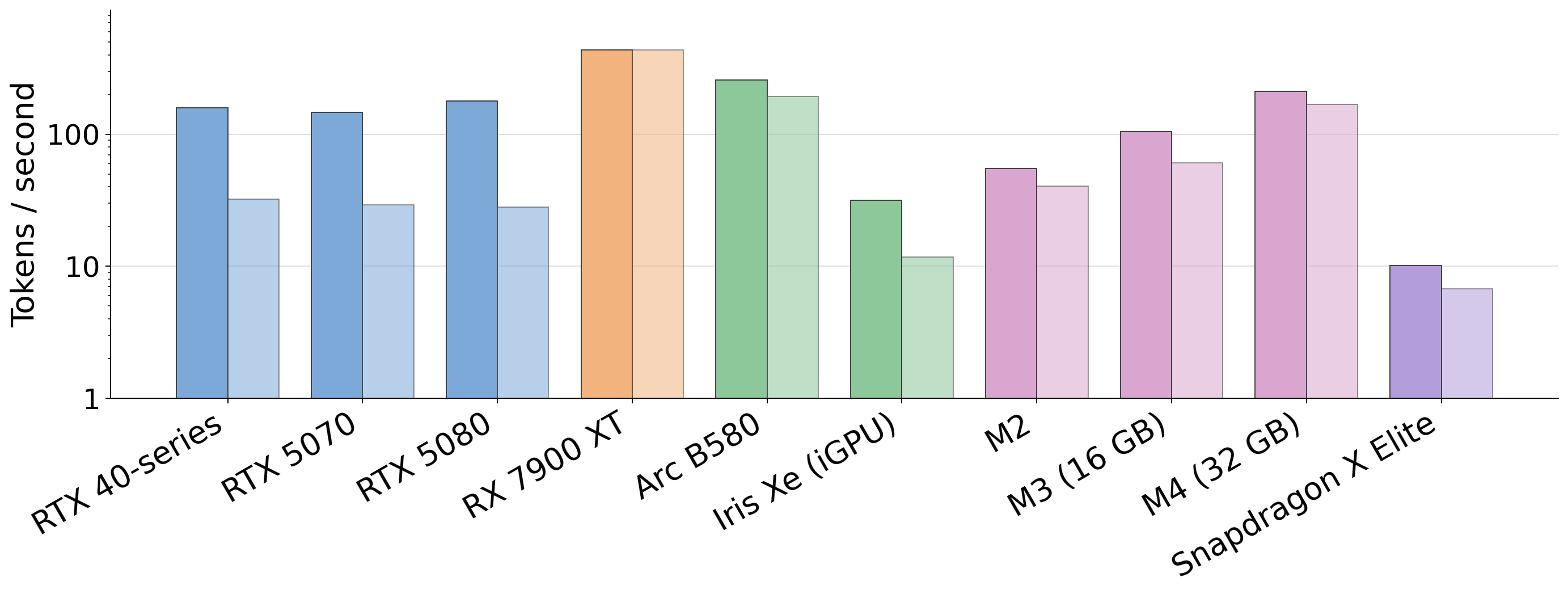}} &
\subfloat[\texttt{smollm} decode]{\includegraphics[width=0.48\textwidth]{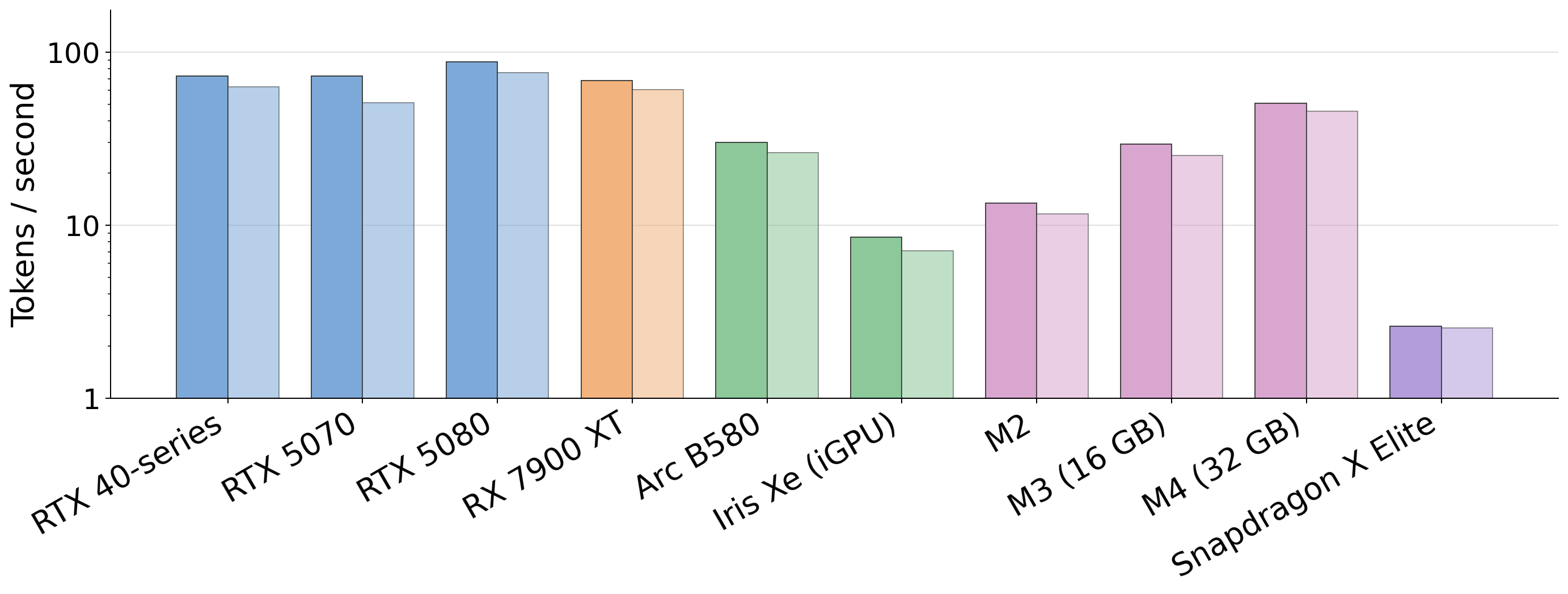}} \\

\subfloat[\texttt{ministral} prefill]{\includegraphics[width=0.48\textwidth]{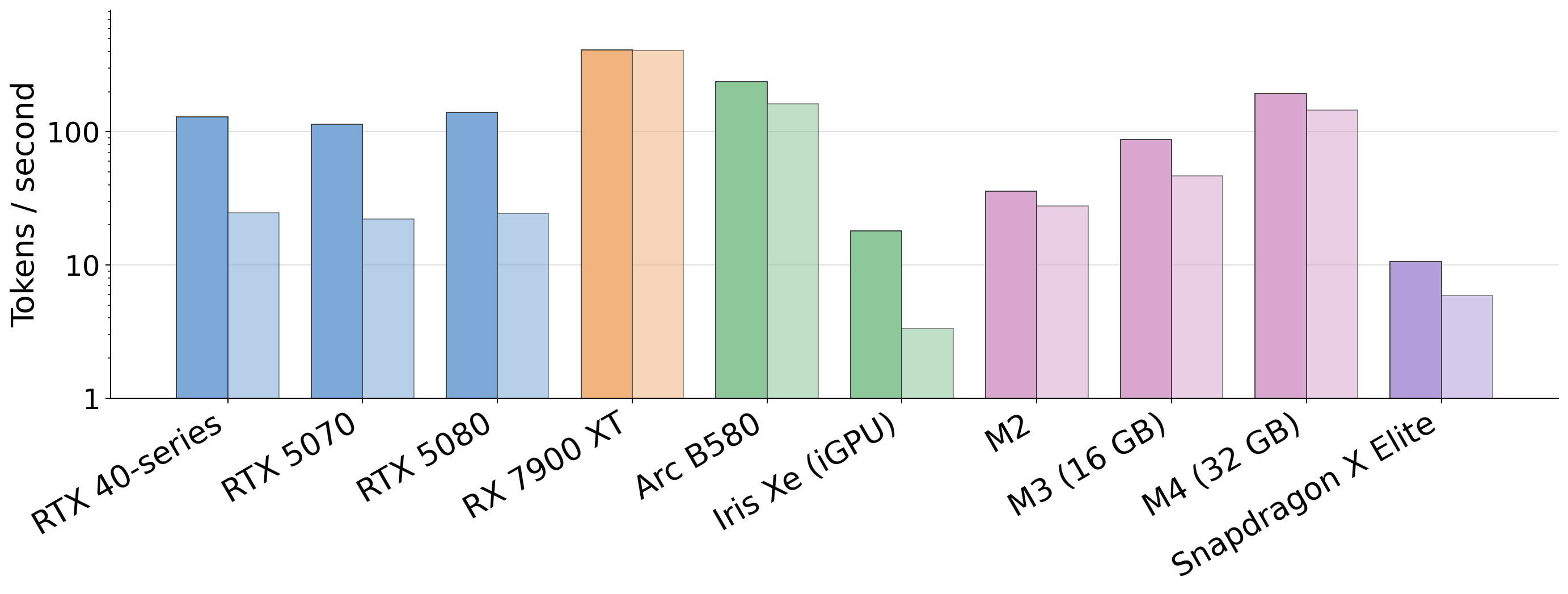}} &
\subfloat[\texttt{ministral} decode]{\includegraphics[width=0.48\textwidth]{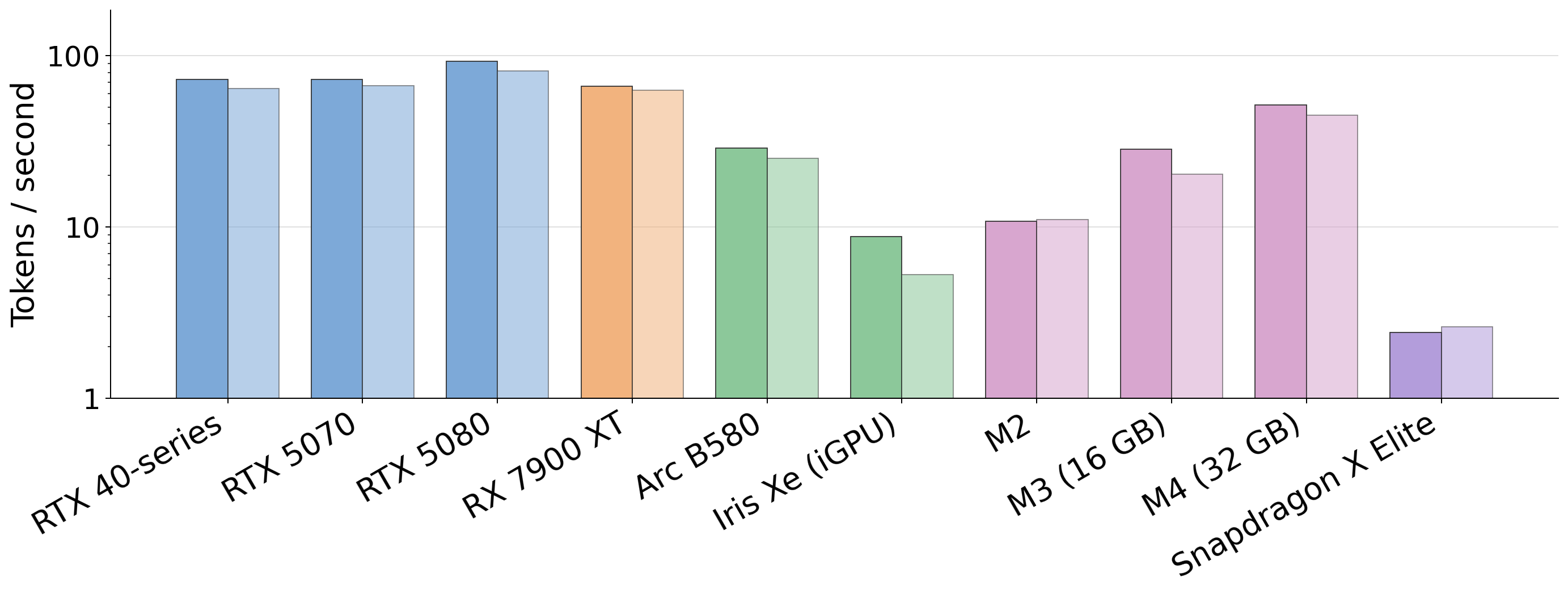}} \\

\subfloat[\texttt{gemma4} prefill]{\includegraphics[width=0.48\textwidth]{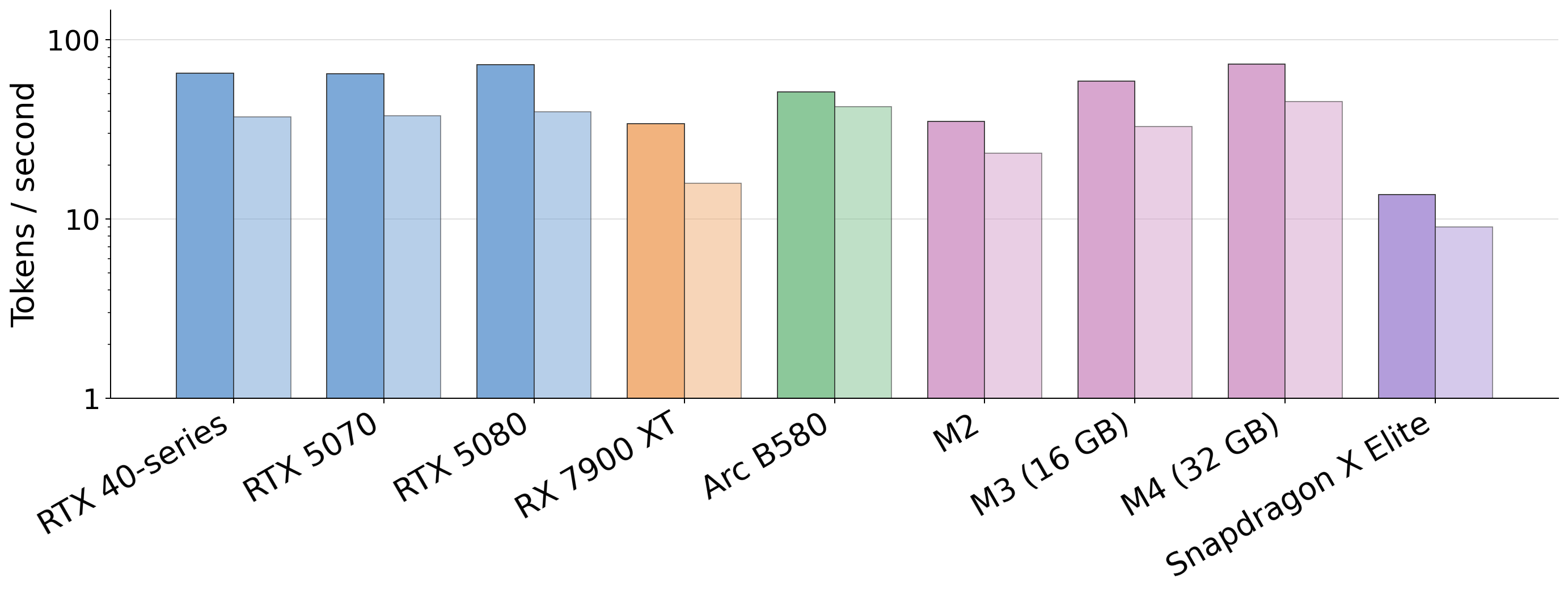}} &
\subfloat[\texttt{gemma4} decode]{\includegraphics[width=0.48\textwidth]{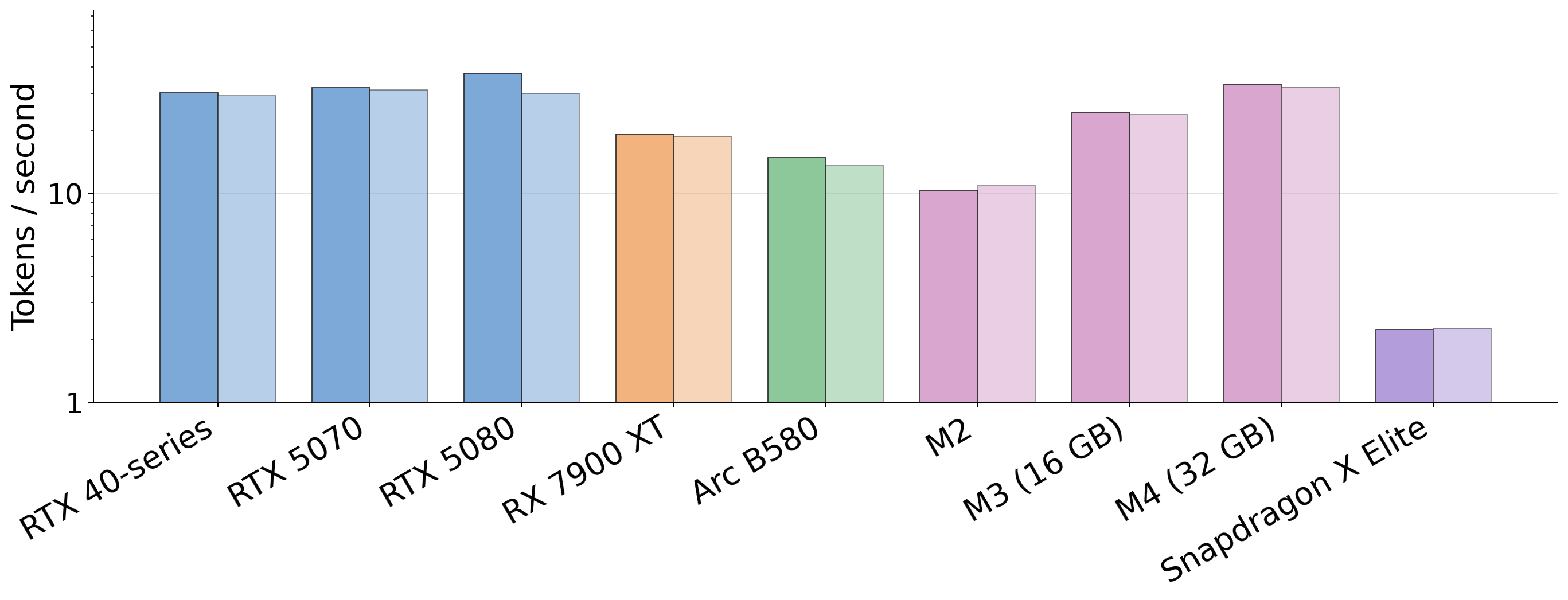}} \\

\subfloat[\texttt{llama} prefill]{\includegraphics[width=0.48\textwidth]{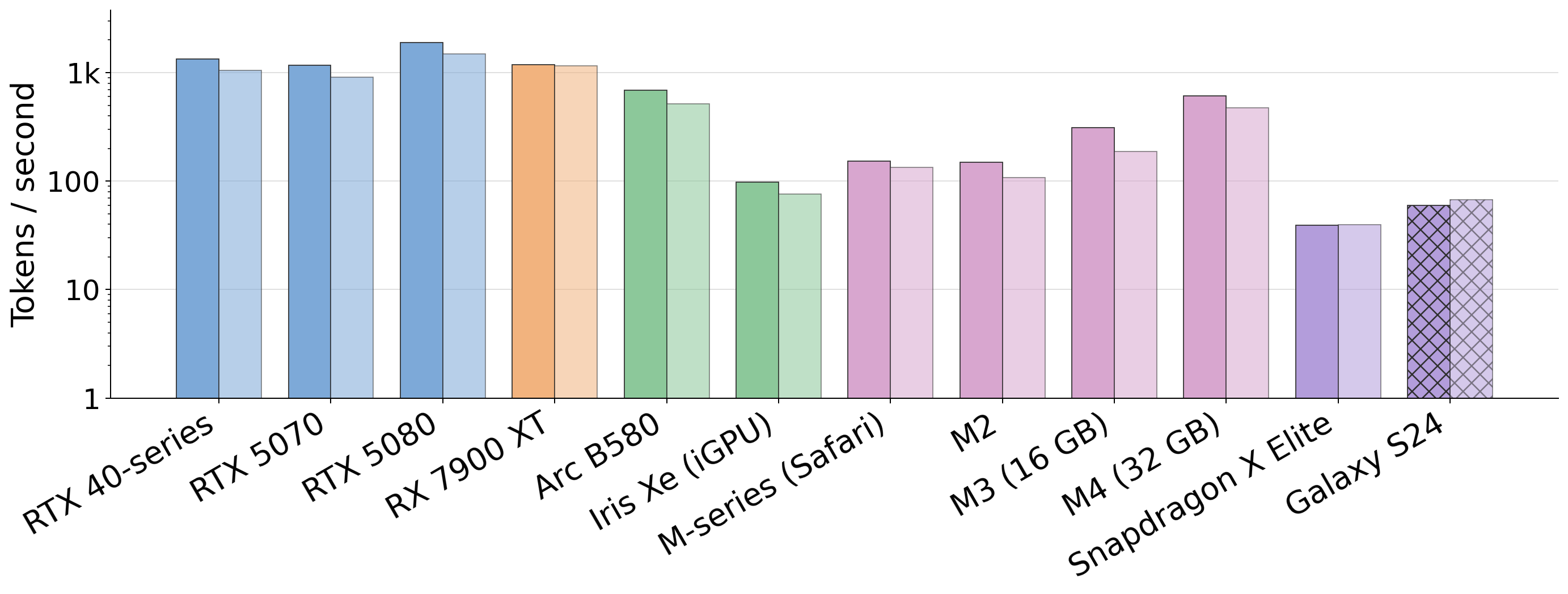}} &
\subfloat[\texttt{llama} decode]{\includegraphics[width=0.48\textwidth]{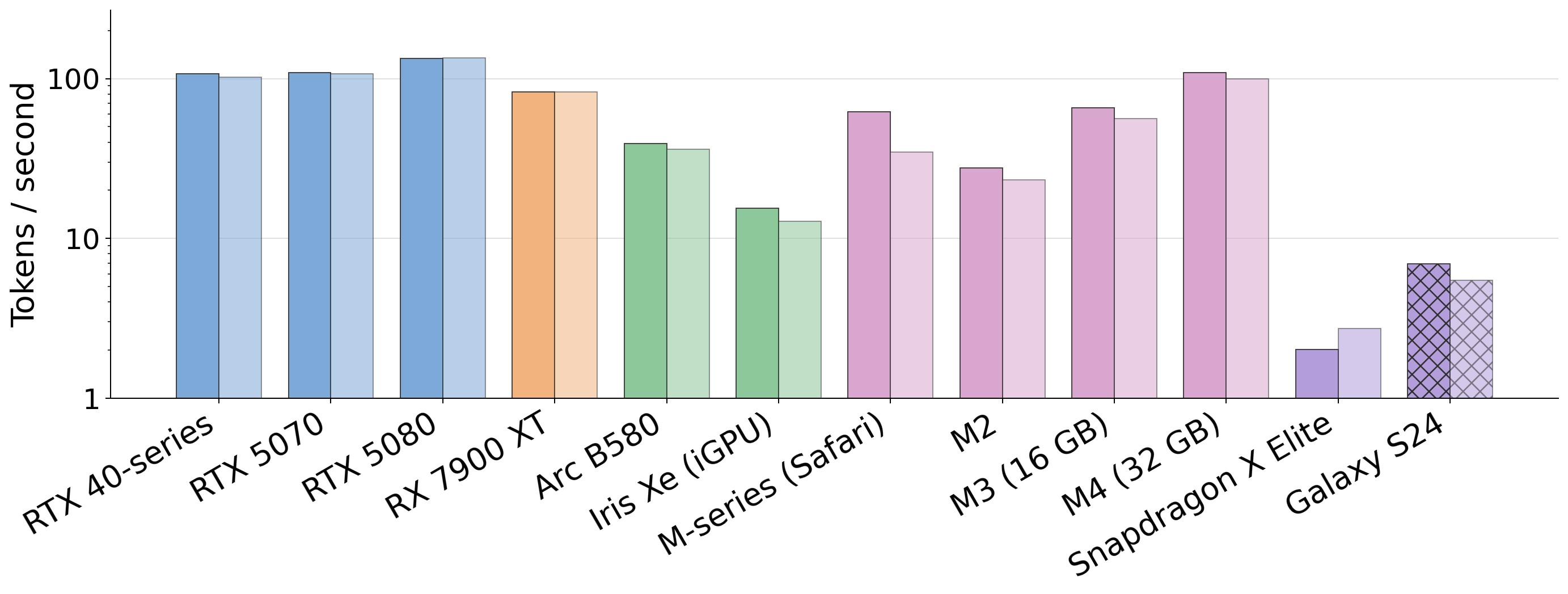}}
\end{tabular}

\caption{Per-device prefill and decode throughput by model, part 2 of 2.}
\label{fig:port-perdev-2}
\end{figure*}

  Each sub-figure in \myfig\ref{fig:port-perdev-1} and \myfig\ref{fig:port-perdev-2} reports prefill and decode throughput in bar graph form for a single model across every device on which it ran. Bars are colored by device family (NVIDIA, AMD, Intel, Apple-Mac, Apple-iOS, Qualcomm, Samsung, ARM, Imagination Technologies), with a hatch pattern distinguishing families that share a color slot. Within each device, paired bars encode KV-cache depth: solid fill for depth 0 and the same color faded for depth 2048. Models are presented in order of parameter count.

On a couple devices, e.g., the Arm Mali and Imagination Technologies PowerVR GPUs, the \texttt{lfm}, \texttt{gemma3}, \texttt{qwen3}, and \texttt{qwen3.5} models ran successfully at a KV-cache depth of 0, but crashed at a KV-cache depth of 2048. In these cases, we include the successfully collected data from the KV-cache of 0 in our clustering and analysis. 


\section{Framework Comparison Prompts}
\label{app:comp-prompts}
These are the prompts we used to measure prefill and decode performance across different WebGPU inference frameworks for our evaluation in \mysec\ref{sec:eval-comp}:
\begin{itemize}
    \item \textbf{Decode:} \textit{Write a story about a turtle}.
    \item \textbf{Prefill:} \textit{Summarize this story about a turtle. In a small pond nestled among the tall reeds of a lush forest, a tiny turtle named Terry lived a simple life. He spent his days swimming in the pond, chasing after the occasional fish, and basking in the warm sun on a rock near the water's edge. Terry was a bit of an oddity among his fellow turtles, and he had a thirst for adventure. He longed to explore the world beyond his pond, to discover new lands, and to meet new creatures. One day, a strong storm rolled in, bringing heavy rain and powerful winds. The pond began to flood, and Terry found himself swept away by the rushing water. He tumbled through the air, his shell rattling against the rocks, and landed with a splash in a nearby stream. As he struggled to swim back to the pond, Terry spotted a small wooden boat drifting downstream. The boat was old and weathered, but it looked sturdy enough to carry him to safety. Without hesitation, Terry scurried into the boat and began to paddle. The journey was long and arduous, but Terry persevered. He navigated through treacherous rapids and dodged snapping fish. Finally, after what seemed like an eternity, he spotted the familiar outline of his home pond. As he emerged from the boat, Terry was greeted by a group of friendly otters, who had been watching him from a distance. They welcomed him with open arms, and Terry was amazed by their kindness and generosity. The otters took Terry under their wing, teaching him how to catch fish and navigate the waters. They showed him the secret spots where the best food could be found, and they introduced him to their friends, a wise old beaver named Benny. Benny, it turned out, was a master of the forest. He had spent years building a network of hidden tunnels and secret passageways, which he used to transport his family and friends to safety during times of drought or flood. Terry was amazed by the complexity of the beaver's underground world and begged Benny to take him on a journey through the forest. Benny, seeing the excitement in Terry's eyes, agreed to take him on a journey. As they explored the forest together, Terry discovered a hidden world of creatures he had never seen before. There were rabbits with bright pink noses, squirrels with fluffy tails, and even a family of field mice who lived in a cozy little burrow. The journey was long and winding, but Terry was thrilled to be a part of Benny's secret world. He realized that there was more to life than just swimming in the pond and catching fish. He had discovered a new passion, one that would take him on many more adventures in the years to come. From that day on, Terry became known as the greatest turtle explorer in the forest. He continued to visit his friends and family, but he also began to explore the world beyond his pond. He discovered hidden waterfalls, secret meadows, and even a hidden cave or two. And though he still lived in his pond, Terry knew that he had found a new home, one that was full of adventure, friendship, and the thrill of discovery.}
\end{itemize}

  \section{Per-Device Quantization Results}
  \label{app:quant-perdev}

\begin{figure*}[p]
\centering
\setlength{\tabcolsep}{4pt}

\begin{tabular}{cc}

\subfloat[Per-device prefill throughput across models.]{
\includegraphics[width=0.47\textwidth]
{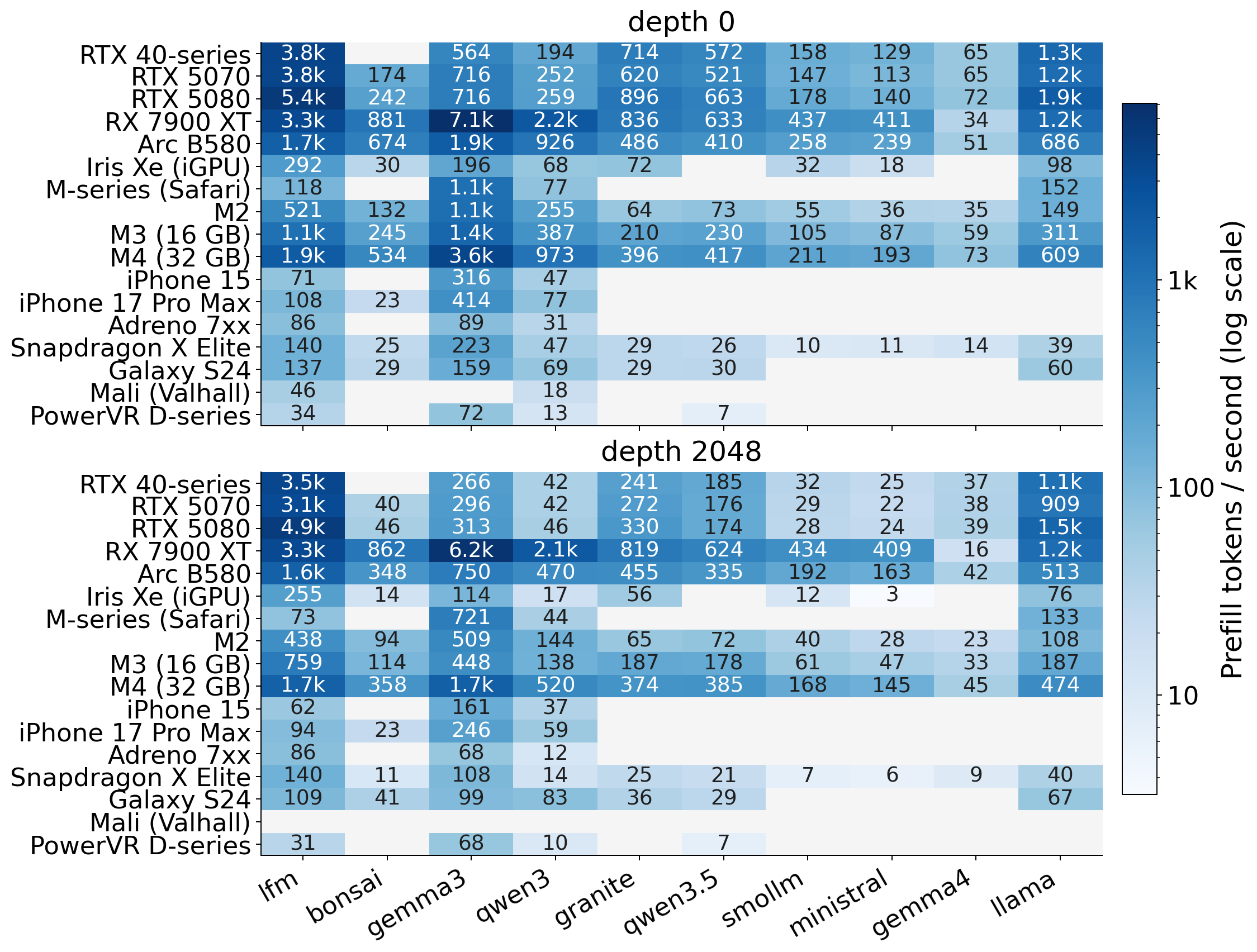}
\label{fig:port-coverage-prefill}
}
&
\subfloat[Per-device decode throughput across models.]{
\includegraphics[width=0.47\textwidth]
{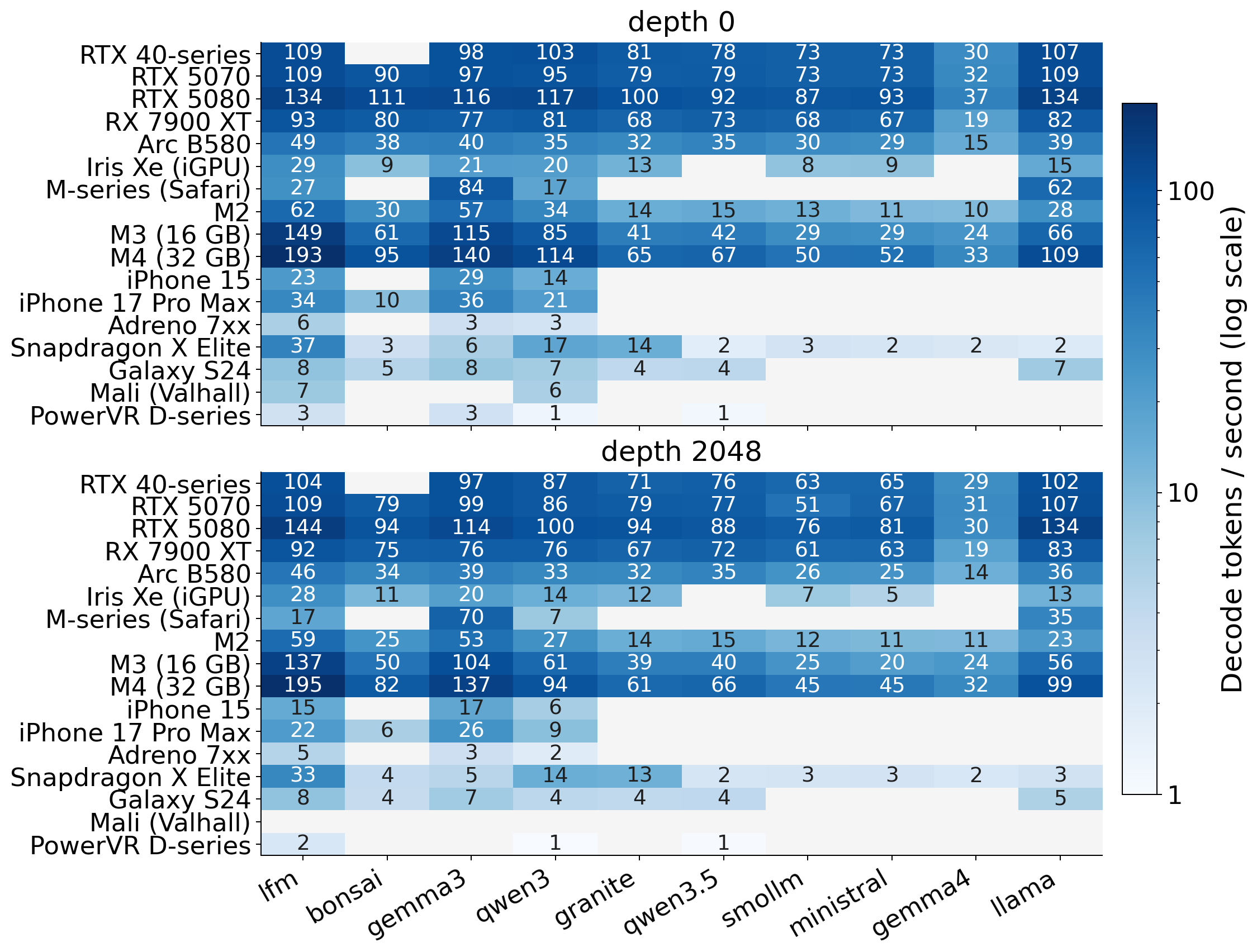}
\label{fig:port-coverage-decode}
}
\\

\subfloat[Per-device prefill throughput across model weight formats.]{
\includegraphics[width=0.47\textwidth]
{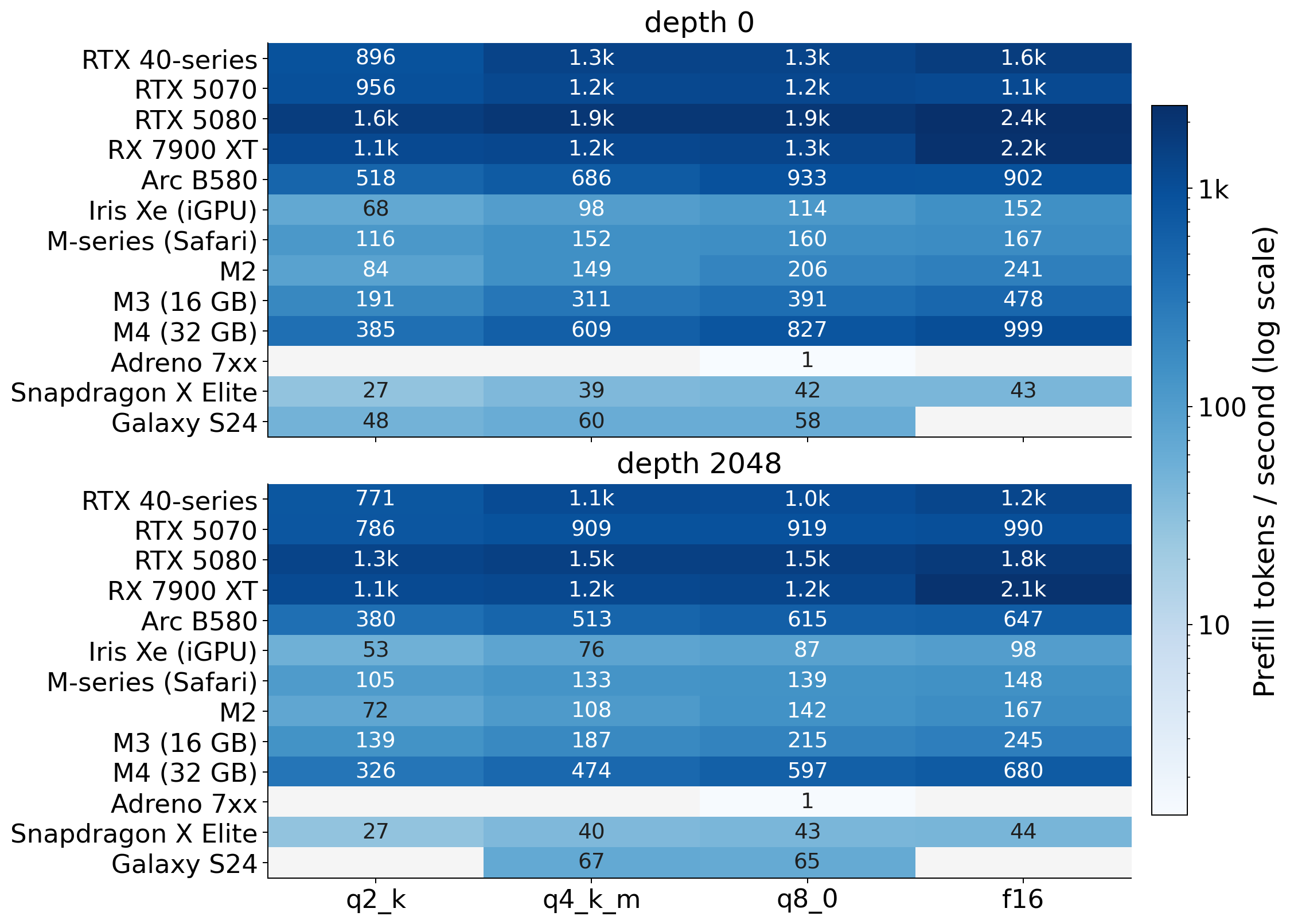}
\label{fig:quant-coverage-prefill}
}
&
\subfloat[Per-device decode throughput across model weight formats.]{
\includegraphics[width=0.47\textwidth]
{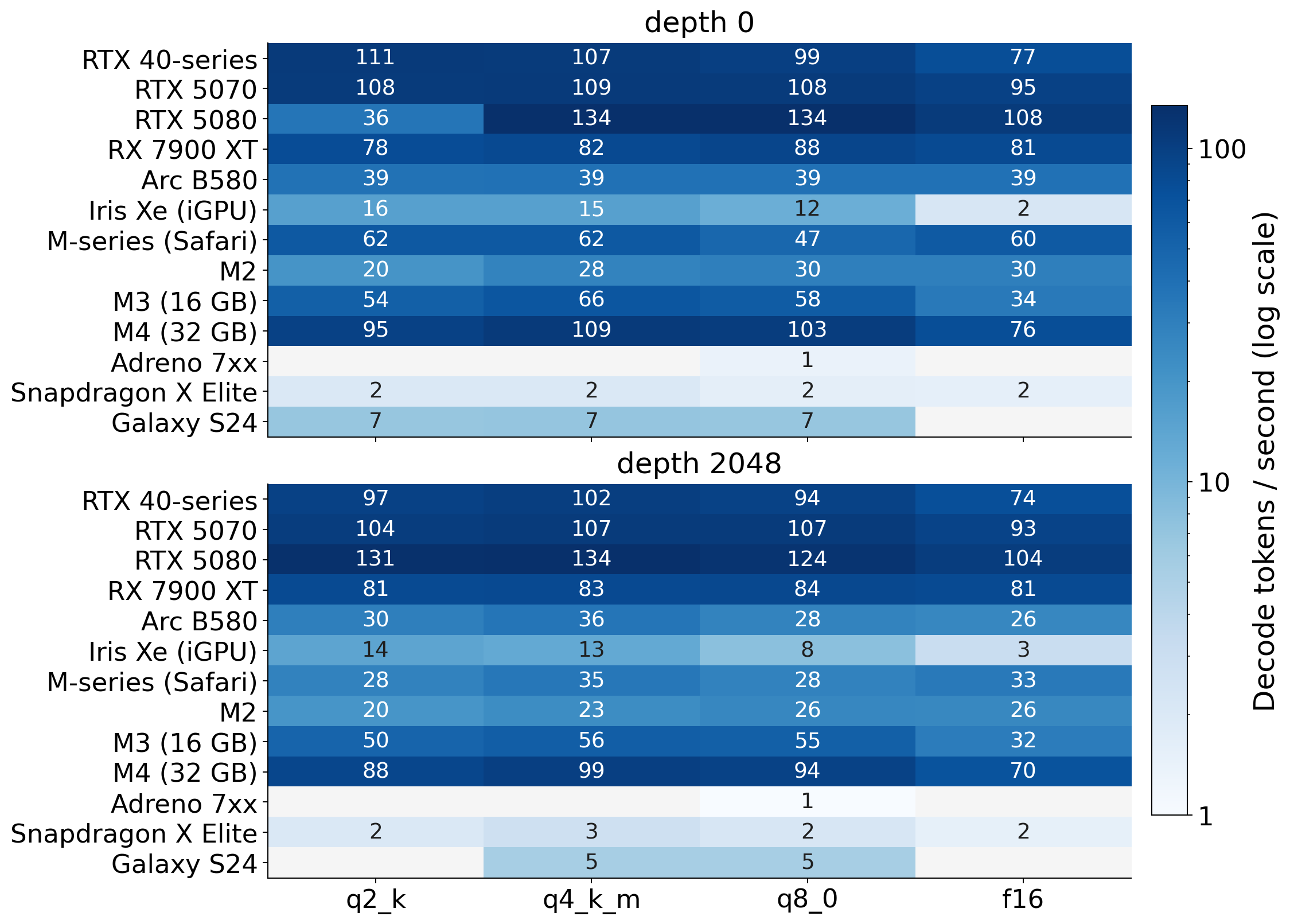}
\label{fig:quant-coverage-decode}
}

\end{tabular}

\caption{
Coverage matrices for the portability and cross-quantization
studies. Each cell reports throughput on a shared log color
scale; blank cells indicate that the model or quantization
variant did not run on that device.
Heatmaps within each panel show KV-cache depth 0
(top) and 2048 (bottom).
}
\label{fig:coverage-combined}
\end{figure*}

\begin{figure*}[t]
\centering
\setlength{\tabcolsep}{2pt}
\renewcommand{\arraystretch}{0.95}

\begin{tabular}{cc}
\subfloat[\texttt{q2\_k} prefill]{
\includegraphics[width=0.48\textwidth]
{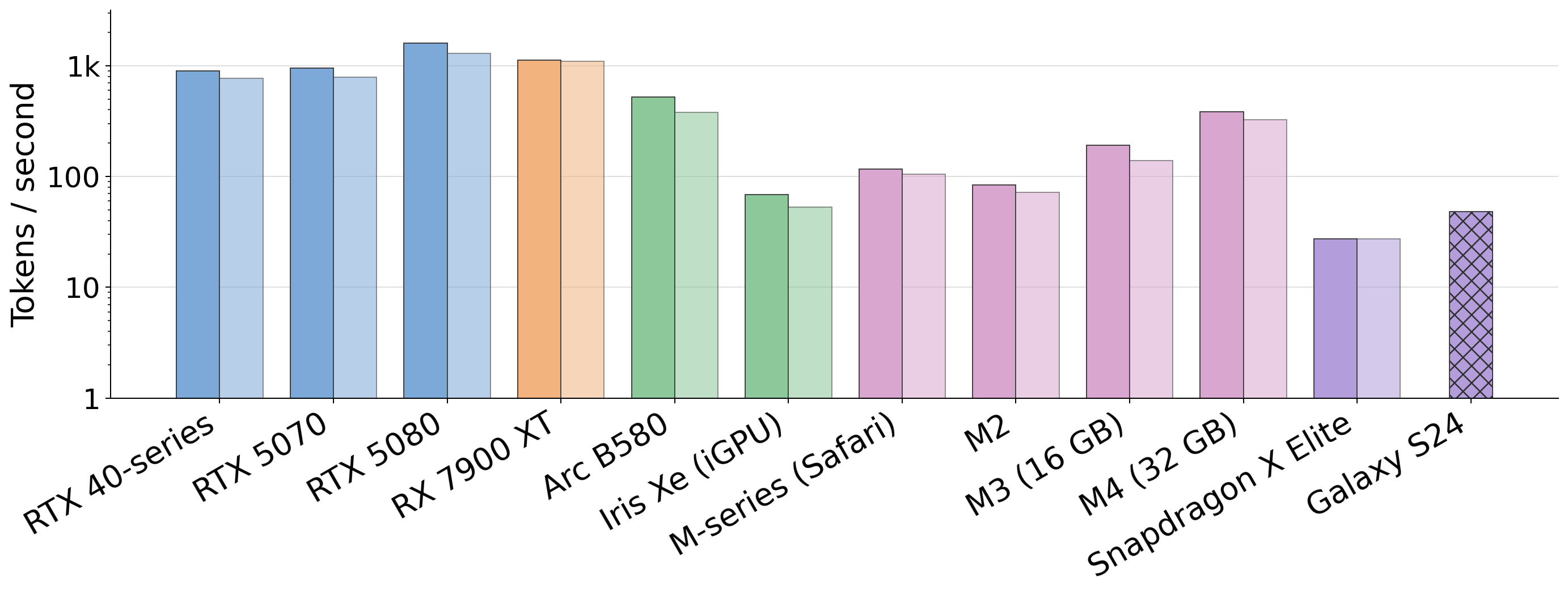}
} &
\subfloat[\texttt{q2\_k} decode]{
\includegraphics[width=0.48\textwidth]
{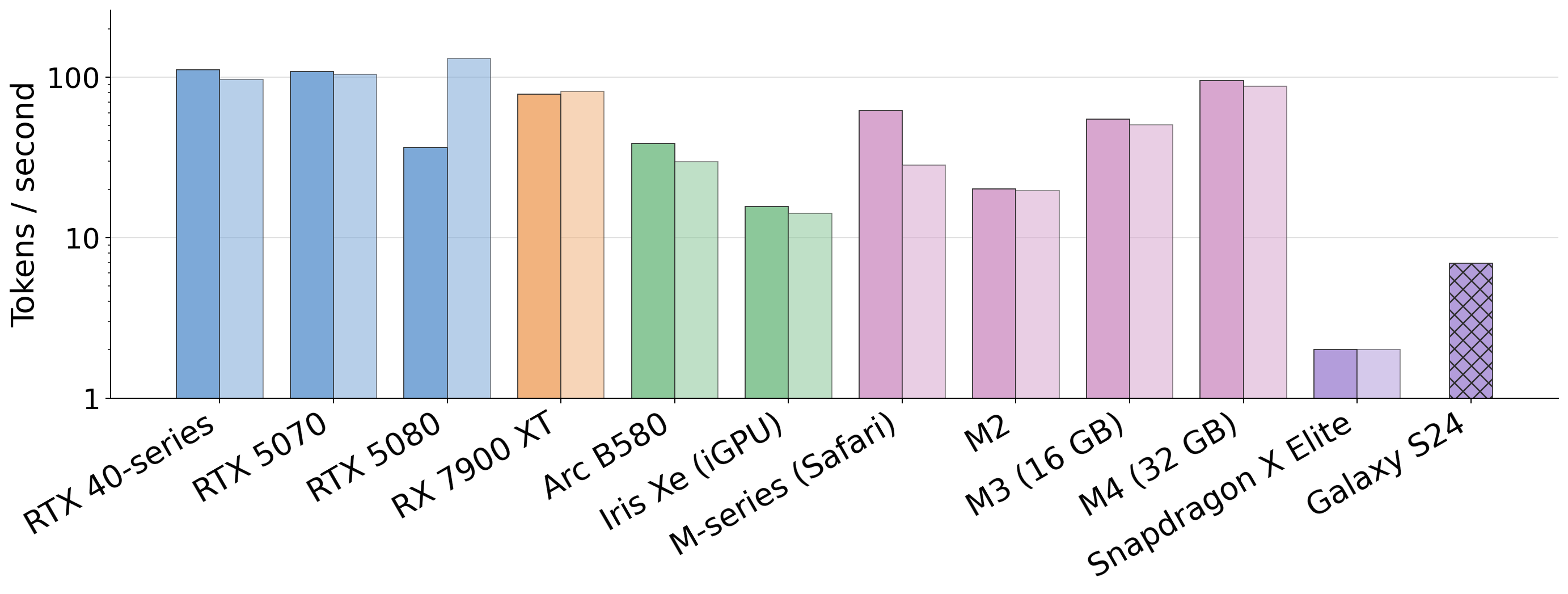}
} \\

\subfloat[\texttt{q4\_k\_m} prefill]{
\includegraphics[width=0.48\textwidth]
{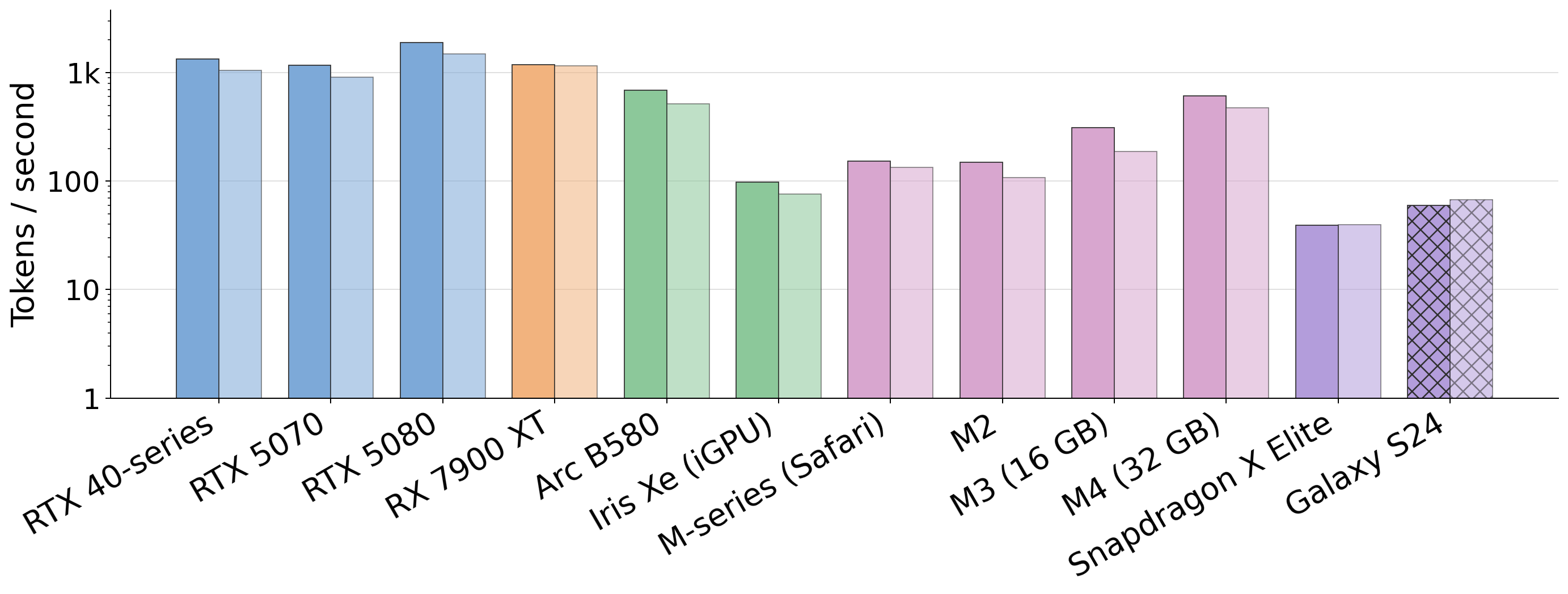}
} &
\subfloat[\texttt{q4\_k\_m} decode]{
\includegraphics[width=0.48\textwidth]
{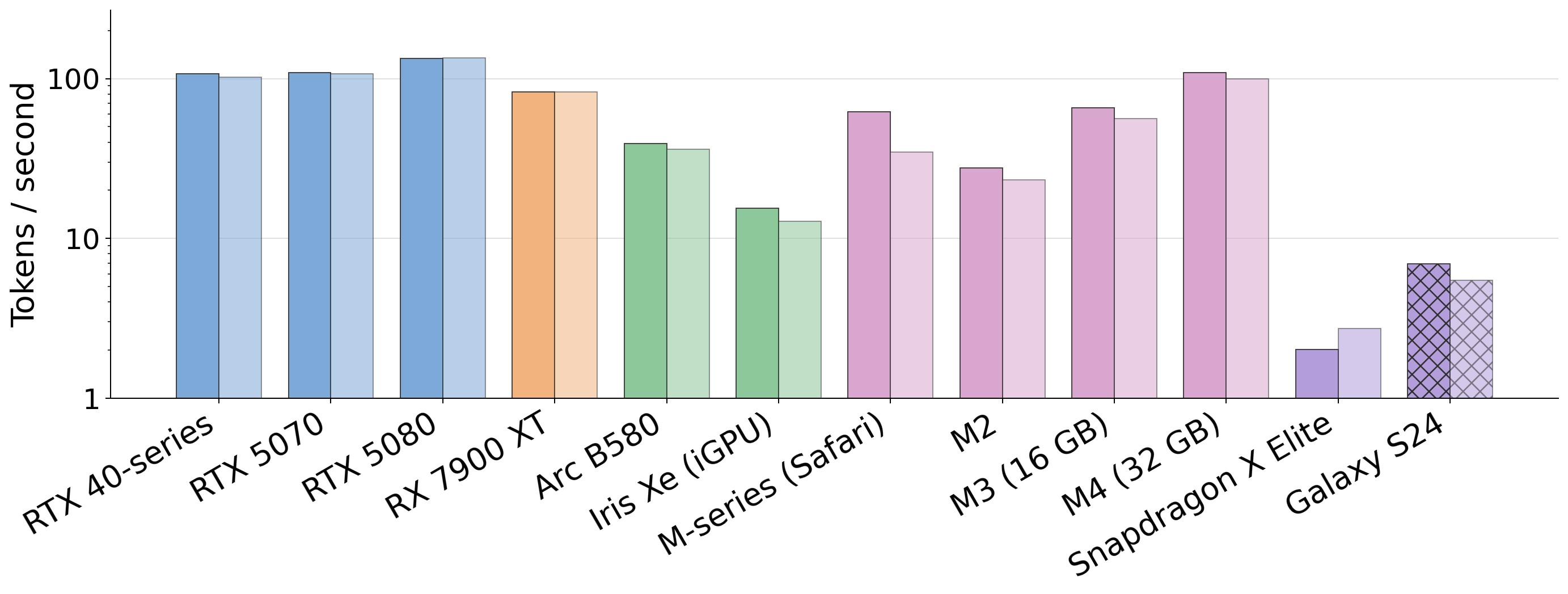}
} \\

\subfloat[\texttt{q8\_0} prefill]{
\includegraphics[width=0.48\textwidth]
{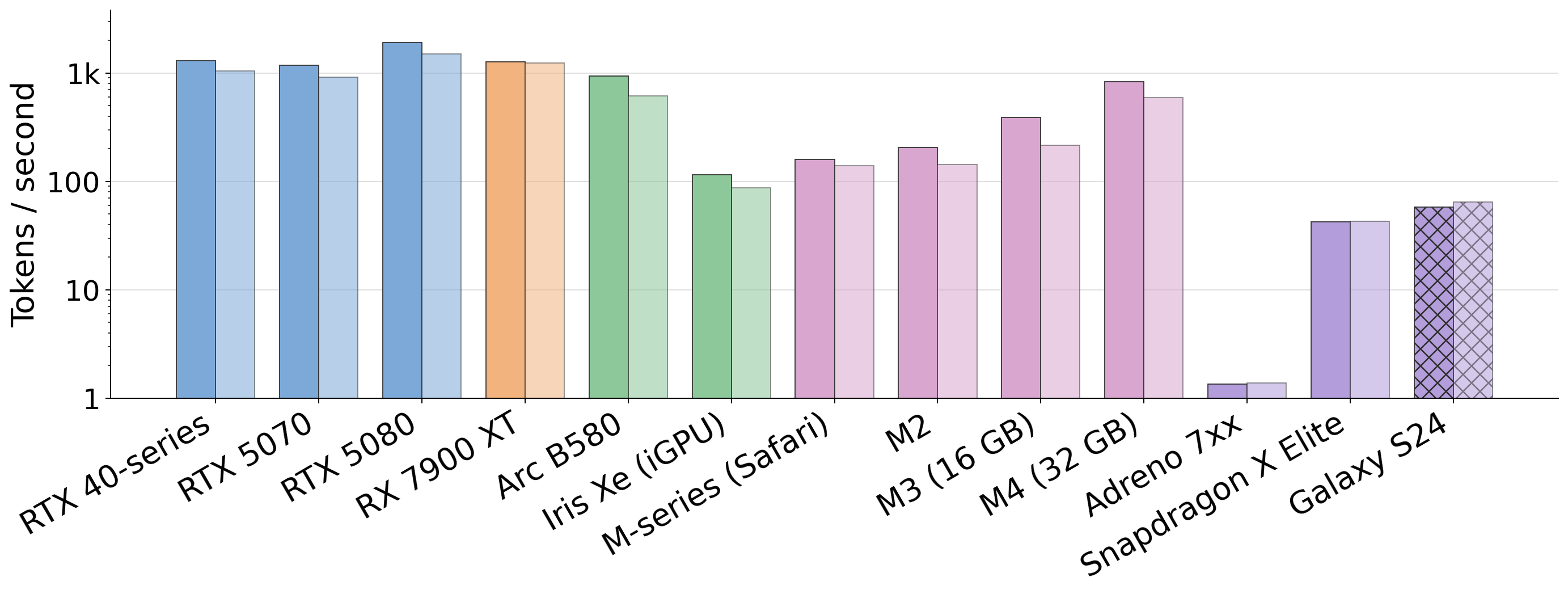}
} &
\subfloat[\texttt{q8\_0} decode]{
\includegraphics[width=0.48\textwidth]
{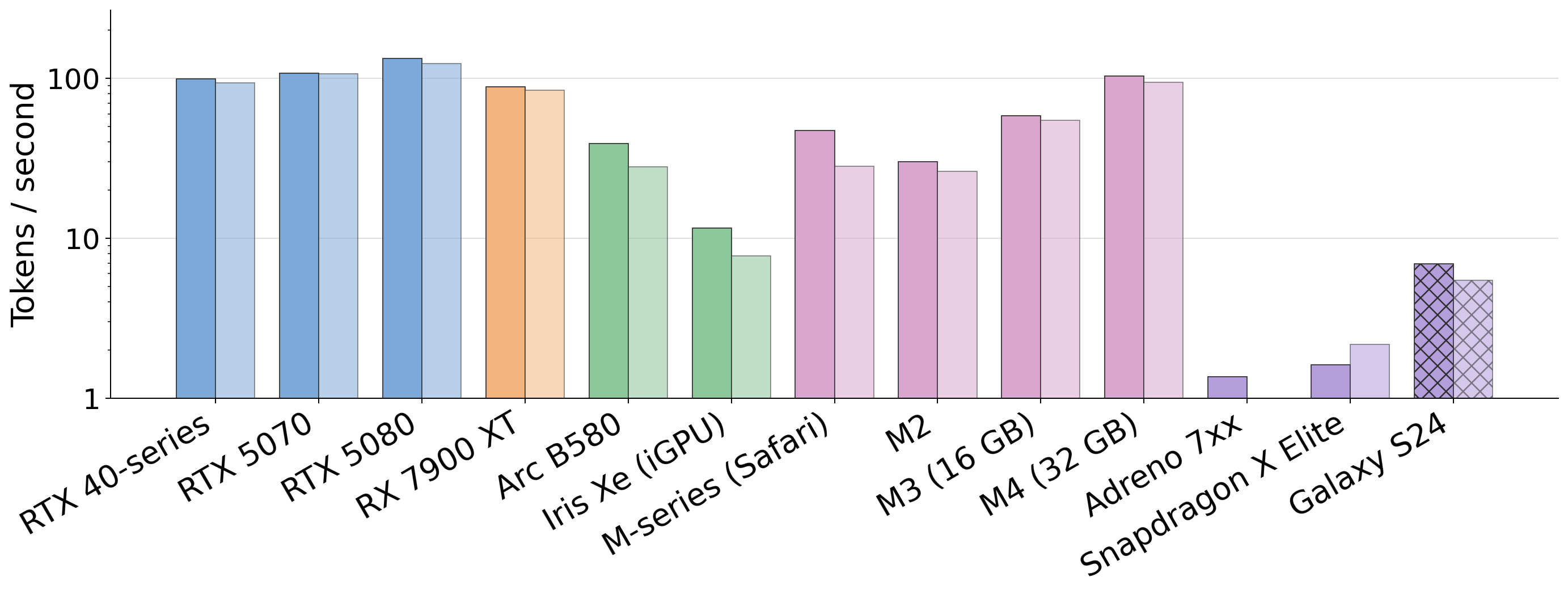}
} \\

\subfloat[\texttt{f16} prefill]{
\includegraphics[width=0.48\textwidth]
{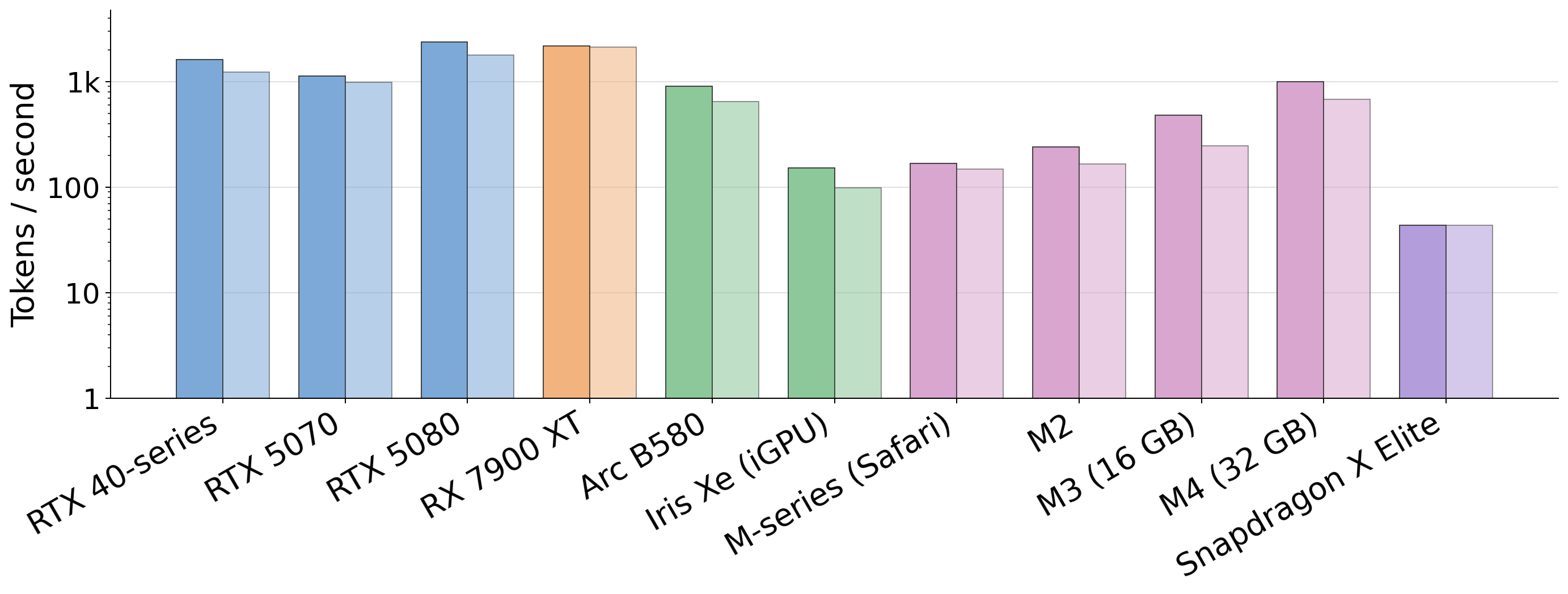}
} &
\subfloat[\texttt{f16} decode]{
\includegraphics[width=0.48\textwidth]
{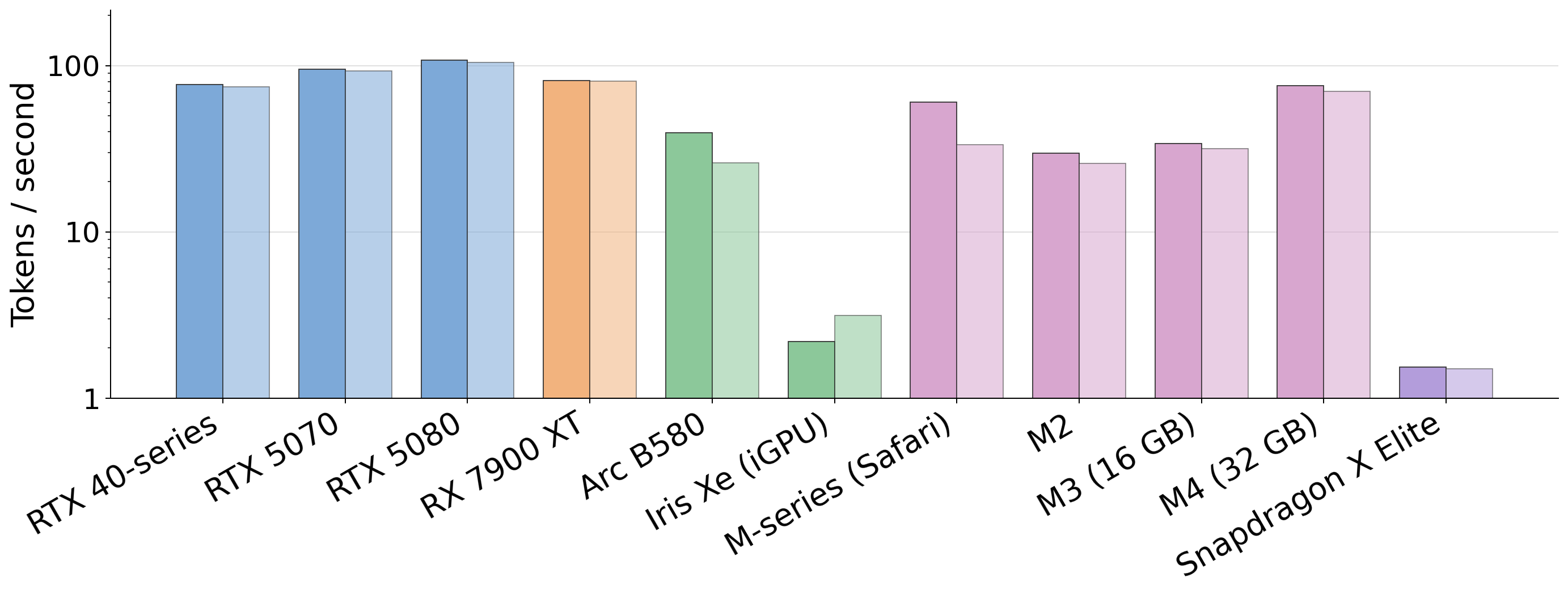}
}
\end{tabular}

\caption{
Per-device prefill and decode throughput for \texttt{llama}
under different model weight formats.
}
\label{fig:quant-perdev}
\end{figure*}

  This appendix supplements \mysec\ref{sec:eval-quant} with full per-device data for the cross-quantization study. Similar to \myapp\ref{app:port-perdev}, \myfig\ref{fig:quant-coverage-prefill} and \myfig\ref{fig:quant-coverage-decode} show the coverage and prefill and decode throughput respectively of running the \texttt{llama} model with different weight formats across the devices in our study, and each sub-figure in \myfig\ref{fig:quant-perdev} reports the throughput per-device and weight format in bar graph form.

\end{document}